\documentstyle[12pt,epsfig]{article}

\begin{document}   

\def\beq{\begin{equation}}
\def\eeq{\end{equation}}
\def\eq{\beq\eeq}
\def\beqn{\begin{eqnarray}}
\def\eeqn{\end{eqnarray}}
\relax
\def\ba{\begin{array}}
\def\ea{\end{array}}
\def\squ{\tilde{Q}}
\def\tb{\mbox{tg$\beta$}}
\def\hH{\hat{H}}
\def\tu{\tilde{u}}
\def\td{\tilde{d}}
\def\te{\tilde{e}}
\def\tQ{\tilde{Q}}
\def\tL{\tilde{L}}
\def\tt{\tilde{t}}
\def\tb{\tilde{b}}
\def\ttau{\tilde{\tau}}
\def\tnu{\tilde{\nu}}
\def\ra{\rightarrow} 
\newcommand{\nn}{\nonumber}
\newcommand{\lsim}{\raisebox{-0.13cm}{~\shortstack{$<$ \\[-0.07cm] $\sim$}}~}
\newcommand{\gsim}{\raisebox{-0.13cm}{~\shortstack{$>$ \\[-0.07cm] $\sim$}}~}
\newcommand{\s}{\\ \vspace*{-2mm} }
\renewcommand{\theequation}{\thesection.\arabic{equation}}
\newcommand{\defprod}{\raisebox{-0.13cm}{~\shortstack{$\prod$ \\[-0.07cm] 
${}_{i=1,4}$}}~}

\begin{flushright}
PM/99--30 \\
hep-ph/9907360
\end{flushright}

\vspace{1cm}

\begin{center}

{\large\sc {\bf 
Phases in the gaugino sector:\\ 
direct reconstruction of the basic parameters \\ 
and impact on the neutralino pair production\\}} 
\vspace{1cm}
{ J.-L. Kneur~\footnote{kneur@lpm.univ-montp2.fr} and 
G. Moultaka~\footnote{moultaka@lpm.univ-montp2.fr}}
\vspace{.5cm}

Physique Math\'ematique et Th\'eorique, UMR No 5825--CNRS, \\
Universit\'e Montpellier II, F--34095 Montpellier Cedex 5, France.

\end{center}

\vspace*{1.5cm}

\begin{abstract}
\noindent 
We consider recovering analytically 
the (generally complex) parameters $\mu$,
$M_1$ and $M_2$ of the gaugino and Higgsino Lagrangian, from 
appropriate physical input in the chargino and neutralino
sectors.  For given $\tan\beta$, we obtain very simple analytic
solutions for $M_2$, $\vert \mu\vert$, $Arg[\mu]$ in the chargino
sector and  a twofold $\vert M_1 \vert$, $Arg[M_1]$ analytic solution
in the neutralino sector, assuming two chargino, two neutralino
masses, and one of the chargino mixing angles as physical input. 
The twofold ambiguity in the neutralino parameters  reconstruction   
may be essentially resolved 
by measuring the $e^+e^- \to  \chi^0_1 \chi^0_2$
production cross-section at future
linear collider energies, which we 
study explicitly with the phase dependences.  
Some salient features and specific
properties of this complex case gaugino "spectrum inversion" are
illustrated and compared with the similar inversion in the real case.
In particular, our algorithms exhibit in a direct and transparent way
the non-trivial theoretical correlation among the chargino and
neutralino parameters, and the resulting allowed domains
when only a subset of the required physical input masses and
production cross-sections is known.

\end{abstract}

\newpage
\setlength{\baselineskip}{15pt}

\section{Introduction}
\setcounter{footnote}{0}
In the
Minimal Supersymmetric extension 
of the Standard Model (MSSM)~\cite{R1, martin}, 
without additional theoretical assumptions, the soft--supersymmetry breaking
part of the Lagrangian\cite{softbreak} involves a large
number of arbitrary parameters. Moreover, many of these parameters may be
in general complex, adding  new sources of
CP violation with respect
to the standard model~\cite{CP1,CPrev}. Apart from 
the many possible phases in the 
flavour sector which are mostly
severely constrained~\cite{fcncexp,PDG98,CPrev},
the other ``flavour-blind" possible phases, appearing 
in particular in the gaugino and/or
Higgs sector, are essentially constrained~\cite{FalOl} by
the electric dipole moment of the electron\cite{edme} and the
neutron\cite{edmn}. As recently
emphasized\cite{IbNa,BrKa},  large phases in this sector are not
excluded at present, although it may be considered as unnatural because
it generally requires specific cancellations among the different
contributing  phases\cite{BartlCP,Savoyetal}. Irrespective of a
realistic theoretical scenario implying such phases, the latter lead, from a 
more pragmatic point of view,  to potentially drastic changes in the
phenomenology of the Higgs and gaugino sectors of the MSSM, therefore
affecting the already challenging reconstruction from 
data~\cite{SNOWMASS}--\cite{choi2} of the structure of the
SUSY and soft-SUSY breaking Lagrangian.  Typically, assuming that a
certain amount of the  SUSY partners may be discovered and some of
their masses, production cross-sections and decay widths measured, 
additional questions arise in the presence of non-zero phases. 
Even without direct measurement of manifestly CP-violating
observables, one may for example try to determine whether a set of
available data necessarily implies the existence of non-zero phases, or
 conversely whether the data is consistent with the underlying MSSM
basic Lagrangian with phases.\\     

In a previous paper~\cite{inoinvreal}, we have given a specific 
strategy and algorithm to  de--diagonalize or ``invert'' 
the parameters in the ino sector, i.e. obtaining some of the
MSSM Lagrangian parameters in direct analytic form in terms of physical
parameters (see also ref. \cite{bartlold} for an early such attempt).
However, our construction was restricted to the
case of real parameters. In the present paper, we
generalize  our procedure to the complex parameter case.  While the
techniques used are quite similar, especially for the less
straightforward 4 $\times$ 4 neutralino mass matrix inversion, the 
choice of the most appropriate input is obviously different, since the
number of input/output parameters is necessarily larger when phases are
present. In addition, some peculiarities of the complex
case require a slightly different approach. As we shall
see, it turns out that the complex case inversion is much more
constrained than one may think at first: strong correlations
between the chargino and neutralino physical masses occur solely
from the structure of the MSSM Lagrangian, despite the additional phase
freedom, and we will illustrate how this somewhat unconventional
spectrum inversion algorithm can exhibit in a more direct and
clearer way such non-trivial correlations. 
For instance, it becomes straightforward in
that way to exhibit allowed domains for the neutralino masses, once
some of the parameters of the chargino sector are determined (or
vice-versa), a task that would usually require rather tedious
systematic scanning over the basic Lagrangian parameters with painstaking
comparisons of physical output results.  

For definiteness we shall only consider here the  
reconstruction of the chargino and neutralino sector parameters with phases, 
letting aside e.g. the possible implications of non trivial phases on
the MSSM Higgs sector phenomenology, which is somehow decoupled from our
analysis and has been
investigated recently in refs. \cite{Hphases,PilWag}. Also, as in ref.
\cite{inoinvreal}, we emphasize that we are not aiming at a new
specific MSSM parameters reconstruction method from experimental
observables. Clearly an algebraic approach alone cannot replace more
complete studies of the reconstruction of the MSSM parameters from data.
Our algorithm should rather be viewed as 
an efficient tool, to complement any more systematic analysis of the data.\\

Our brute algorithm 
gives analytical solutions in the {\em unconstrained} MSSM (i.e. no
universality of the gaugino mass terms is assumed), for $\mu \equiv
\vert\mu\vert e^{i \Phi_\mu}$, $M_2$, and $M_1 \equiv \vert M_1\vert
e^{i \Phi_{M_1}}$, using as input two chargino masses $M_{\chi_1}$,
$M_{\chi_2}$, two (arbitrary) neutralino masses $M_{N_i}$, $M_{N_j}$,
and one chargino mixing angle, plus $\tan\beta$~\footnote{This
algorithm is available
as a fortran code upon request to the authors.}. As will be explained, 
we can choose a phase convention such that $M_2$ is real, 
so that the counting of physical
input versus output parameters is consistent.  
Obviously, any of the
latter input, if not known, may be scanned over some specified range 
to determine e.g. the resulting allowed domain in the output parameters
$\mu$, $M_1$, $M_2$.
As a by-product, we obtain the neutralino diagonalization
matrix elements --and therefore the neutralino physical couplings - as
direct functions of the masses, and we have recalculated the $e^+e^- \to 
\chi^0_i \chi^0_j $ production cross-section in this framework with 
complex $\mu$, $M_1$. 
The neutralino 1,2 pair production cross-section has been widely
studied as a very promising detection process at 
LEP2~\cite{ambrosanio,LEPYR96,inobound} 
or at a future $e^+e^-$ linear collider~\cite{lcrep,n1n2lc}. The next-
to-lightest $M_{N_2}$ mass could be measured 
with good accuracy, and the lightest (LSP)
$M_{N_1}$ as well from chargino decay product~\cite{lcrep}. 
As we shall illustrate,
the neutralino pair production measurement should also
play an important role in
resolving an intrinsic ambiguity in the reconstruction of the
$M_1$ parameter from our algorithm.
One should note, however, that prior to our analysis,
the precise reconstruction of the chargino masses and mixing
angles from actual data will be non trivial. This issue was recently
investigated in details in ref. \cite{choi1,choi2}, with another way
of reconstructing the $\mu$ and $M_2$  parameters by considering the
chargino pairs which can be produced at a future $e^+e^-$ linear
collider and studying theirs spin correlation properties. Accordingly,
our analysis is very complementary to theirs, in particular for the
neutralino sector.\\ 
In section 2, we  briefly recall the general
parameterization of the gaugino and Higgsino sector, and the physically
relevant phases fixing our  conventions.  In section 3, we  discuss the
complex chargino and neutralino de--diagonalization algorithm and
discuss uniqueness conditions.  Typical illustrations of the
reconstruction  of the parameters are given, and in particular 
the kind of non-trivial correlations 
that one can glean in the case of partial knowledge of the input. 
In section 4 we give analytical expressions for the 
$e^+e^- \to  \chi^0_i \chi^0_j $ production cross-section in this
general complex $\mu$, $M_1$ case and illustrate how it can resolve
the two-fold ambiguity for $M_1$, when determined from the neutralino masses 
alone. 
Finally  section 5 gives some conclusions and outlook.     

\section{Phases in the MSSM gaugino Lagrangian}
\setcounter{equation}{0}
The relevant Gaugino/Higgsino part of the unconstrained MSSM Lagrangian 
can be found in many different places, see e.g. refs\cite{R1,martin}.
More specifically 
the soft SUSY-breaking Majorana mass terms: \\
\beq
\label{Lgaugino}
{\cal L}_{gaugino} = -\frac{M_1}{2} \tilde{B} \tilde{B} 
-\frac{M_2}{2} \tilde{W}^i \tilde{W}_i
-\frac{M_3}{2} \tilde{G}^a \tilde{G}_a \;+h.c.
\eeq
have to be supplemented by the supersymmetric $\mu$ term originating
from the quadratic part of the superpotential and contributing to the 
Higgsino terms. Writing explicitly
 for illustration only the terms contributing to neutralinos
(after electroweak symmetry breaking): 

\begin{eqnarray}
\label{Lneutralino}
{\cal L}_{neutralino}&=&  
 m_Z c_w \sin \beta \tilde{W}_3 \tilde{H}_u
- m_Z c_w \cos \beta \tilde{W}_3 \tilde{H}_d \nonumber \\
&& + m_Z s_w \cos \beta \tilde{B} \tilde{H}_d 
   - m_Z s_w \sin \beta \tilde{B} \tilde{H}_u +\mu \tilde{H}_u \tilde{H}_d
\;+h.c.
\nonumber \\ 
\end{eqnarray}
with $s_w \equiv \sin\theta_W$, $c_w \equiv \cos\theta_W$, and
$\tan\beta \equiv v_u/v_d$ the ratio of the two Higgs vacuum
expectation values, $v_{u,d} = \langle H_{u,d} \rangle$, 
it is now easy to see from Eqs.(\ref{Lgaugino}, \ref{Lneutralino}) that
only the relative phases of say $(M_2, M_1)$ and $(M_2 , \mu)$ are physically
relevant.  
Indeed any phase change of $M_2$ in Eq.(\ref{Lgaugino}) can always be
absorbed by a phase change of the $\tilde{W}$ field. The latter, however,
fixes uniquely the phase change of $\tilde{H}_d$, $\tilde{H}_u$
and $\tilde{B}$ in Eq.(\ref{Lneutralino}) since we assume all vev's
and gauge couplings to be real, in such a way that
the phases of the combinations $M_2/M_1$ and $\mu M_2^2 $ remain unchanged. 
We can thus choose, without loss of generality, 
$M_2 \geq 0 $ and real, and $M_1$ and $\mu$ to have arbitrary 
phases, which fixes our phase convention. Even though no symmetry arguments
were invoked here for ${\cal L}_{neutralino}$, one should keep in mind
that Peccei-Quinn and R--symmetries are implicitly needed to assure that
phase rotations of the fields, necessary for a given phase convention,
preserve automatically the reality of quantities appearing in the 
PQ- and R-symmetry preserving sector
of the MSSM, such as quark and lepton 
masses and gauge couplings.\footnote{Note that the assumption of no phases
for the Higgs vev's is the only convention here that affects the phases
of the remaining soft susy sectors of the MSSM. For a clear discussion 
on the relevant MSSM phases from $U(1)$ symmetry arguments see, e.g.,
refs.~\cite{PilWag,Savoyetal}.}

\section{Extracting the parameters $\mu$, $M_2$ and $M_1$}
\subsection{Chargino sector}
\setcounter{equation}{0}

In the chargino sector, and with the phase convention discussed above, the
mixing mass matrix reads from Eqs.~(\ref{Lgaugino}, \ref{Lneutralino}) 
\beq
\label{Mchargino}
M_C = \left(
  \begin{array}{cc} M_2  & \sqrt 2 m_W \sin\beta  \\
              \sqrt 2 m_W \cos\beta  & \vert \mu \vert e^{i \Phi_\mu}  
\end{array} 
\right)
\eeq 
The squared chargino masses  
are obtained as the eigenvalues of the
$M^\dagger_C M_C $ matrix:
\beq
\label{Mchi12}
M^2_{\chi_{1,2}} = \frac{1}{2} [ M^2_2 +\vert\mu\vert^2 +2 m^2_W \mp 
\Delta\;]
\eeq
where ($\Delta \equiv M^2_{\chi_2} -M^2_{\chi_1}$) 
\beq
\Delta =\sqrt{(M^2_2-\vert\mu\vert^2)^2 +4m^4_W \cos^2 2\beta 
+4m^2_W(M^2_2+\vert\mu\vert^2) +8 m^2_W
M_2 \vert\mu\vert \sin 2\beta \cos \Phi_\mu}\;.
\eeq
Note that the diagonalization of the non-symmetric matrix (\ref{Mchargino})
is performed via two unitary matrices, involving mixing angles 
$\phi_{L,R}$\footnote{For convenience we adopt notations similar to
those of ref. \cite{choi2}. Note, however, an overall difference of
signs of $\phi_L$, $\phi_R$ angles in eqs. (\ref{phil}), which should be
due to a different
convention for the ordering of the two chargino mass eigenvalues. Our
convention is such that $ M^\dagger_C M_C =
diag(M^2_{\chi_2}, M^2_{\chi_1})$ (with $M^2_{\chi_2} > 
M^2_{\chi_1}$).}:
\beq
U^*_L M_C U^{-1}_R = M_{C,\; diag}
\eeq
with
\beqn
U_L & = &
\left(
  \begin{array}{cc} 
\cos\phi_L & e^{-i\beta_L} \sin\phi_L \\
-e^{i \beta_L} \sin\phi_L & \cos\phi_L \end{array} \right) \\ \nn
&  &   \\ \nn
U_R & = &
\left(
  \begin{array}{cc} 
 e^{i\gamma_1} & 0 \\
0 & e^{i\gamma_2} 
\end{array}\right) 
\left(
  \begin{array}{cc}  
\cos\phi_R & e^{-i\beta_R} \sin\phi_R \\
-e^{i \beta_R} \sin\phi_R & \cos\phi_R \end{array}\right)
\eeqn
where
the phases $\beta_{L,R}$ and $\gamma_{1,2}$ depend on the parameters
$M_2$, $|\mu|$, $\Phi_\mu$ (and $\tan\beta$). Their
explicit expressions, that we 
shall not need in our subsequent analysis,
may be found e.g. in
refs. \cite{choi2,IbNa}.\\ 
From the relations $U^*_L M_C M^\dagger_C
(U^*_L)^{-1} = U_R M^\dagger_C M_C U^{-1}_R = diag(M^2_{\chi_2},
M^2_{\chi_1})$ one obtains 
\beqn
\label{phil} 
\cos 2\phi_L & = & \frac{M^2_2 -\vert\mu\vert^2 -2 m^2_W \cos 2\beta}
{\Delta}  \\ \nn
\sin 2 \phi_L & = & \frac{2 m_W \sqrt{M^2_2 +\vert\mu\vert^2 +
(M^2_2 -\vert\mu\vert^2) \cos 2 \beta +2 M_2 \vert\mu\vert \sin 2\beta
\cos \Phi_\mu } }{\Delta}
\eeqn
and $\phi_R$ obtained by similar expressions in which $\sin\beta
\leftrightarrow \cos\beta$.  \\
Accordingly, in the complex case 
there are two independent parameters among $\phi_L$, $\phi_R$ and $\tan\beta$,
in addition to the two physical chargino masses.
The mixing angles 
$\phi_L$ and $\phi_R$ enter in different combinations
in the $Z\chi^+_i\chi^-_j$ and $e\chi^-_i \nu_e$ couplings, such that,
as worked out in details in ref. \cite{choi1,choi2}, precise
measurements of the $e^+e^- \to \chi^+_i\chi^-_j$ cross-section and
final state polarization properties should allow a 
non-ambiguous reconstruction of the two chargino masses,
mixing angles $\phi_L$ and $\phi_R$, and even $\tan\beta$.
We refer
to their analysis for further details. For convenience, 
in our subsequent analysis, 
we rather choose to have $\tan\beta$
as input, having in mind that $\tan\beta$ maybe also determined
from another sector.\\

Therefore, assuming as input $\tan\beta$, 
the two chargino masses and one mixing angle (say $\phi_L$ for
definiteness), one can derive straightforwardly from eqs. (\ref{phil})  
``inverted" expressions for $\vert\mu\vert$, $\Phi_\mu$ and $M_2$
directly in terms of physical parameters.  These read
\beq
\label{mum2}
|\mu| (M_2) = \left[\frac{1}{2}(\Sigma -2 m^2_W (1 +(-) \cos 2\beta
-(+) \Delta \cos(2\phi_L) )\right]^{1/2}
\eeq
(where $\Sigma \equiv 
M^2_{\chi_2} + M^2_{\chi_1}$) and
\beqn
\label{cphmu}
& & \cos(\Phi_\mu) = 1 -{\frac{M^2_{\chi_1} M^2_{\chi_2}
-(P - m^2_W\sin 2\beta)^2}
{2 m^2_W \: P \;\sin 2 \beta } }\; ; \\ \nn
& & 
P \equiv M_2 \: |\mu| = \frac{1}{2}\:\left[ (\Delta \cos 2 \phi_L +2
m^2_W \cos 2\beta)^2 - (\Sigma -2m^2_W)^2 \right]^{1/2}
\eeqn
Actually, since only $\cos(\Phi_\mu)$ enters the mass formulas
(\ref{Mchi12}), $\Phi_\mu$
is determined by (\ref{cphmu}) up to a twofold ambiguity: $\Phi_\mu
\leftrightarrow 2\pi -\Phi_\mu$. To resolve it requires {\sl a priori} 
the measurement of manifestly CP violating
observables\cite{choi2}. Note however that the only effect of this
ambiguity on our chargino, neutralino inversion algorithm is accordingly
$\Phi_{M_1} \leftrightarrow -\Phi_{M_1}$, without affecting $|\mu|$,
$|M_1|$ or the $e^+e^- \to \chi^0_i \chi^0_j$ production
cross-section, as will be illustrated in subsequent sections.\\

Now, what is not immediately transparent from eqs. (\ref{mum2}, \ref{cphmu}) 
is that they already exhibit relatively strong constraints among the
physical parameters.  Namely, {\em arbitrary} values of 
$M^2_{\chi_2}$, $M^2_{\chi_1}$ and $0 < \phi_L < \pi/2$, are not at all
guaranteed to be consistent with the obvious constraints $M_2, |\mu| \geq 0$, 
and $| \cos\Phi_\mu | \leq 1$. More precisely, one arrives, after some 
straightforward algebra, to the two consistency constraints
 
\beqn
& &  -\left( \Delta^2 \cos^2(2\phi_L) + 
4m^2_W \Delta \cos 2\beta \cos(2\phi_L) +        
4 m^2_W \Sigma - \Delta^2 - 8 m^4_W \sin^2(2\beta) \right)^2 \nn \\ 
& & \leq    
   16 m^4_W \sin^2(2\beta)\: (\Delta \cos(2\phi_L) + 2 m^2_W (1 +
\cos2\beta) - \Sigma) \nn \\  
& & \times \; (\Delta \cos(2\phi_L)  -2 m^2_W (1 -\cos2\beta)
+\Sigma) \;\; \leq \;0  \nn \\
&&  \nn \\
&& \Delta \cos(2\phi_L) + 2 m^2_W (1 +
\cos2\beta) - \Sigma \; \leq \; 0 \nn \\
\label{cphLbound}
\eeqn

In other words, only those values of $M_{\chi_1}$, $M_{\chi_2}$ and $\cos 2\phi_L$ which satisfy 
eqs. (\ref{cphLbound}) are consistent within the MSSM, 
which for some masses may give rather restricted
ranges for $\cos 2\phi_L$. These algebraic constraints mean that,
given $|\mu|$ and $M_2$, one cannot reach arbitrary values for the chargino 
masses despite the phase freedom, the magnitudes of these masses being
essentially determined by $|\mu|$ and $M_2$.   
Obviously, the actual chargino pair production will correspond to only
one consistent $\phi_L$ value, but (\ref{cphLbound}) also gives in a
straightforward way  some other useful information, like for instance
the allowed $(M_{\chi_2}, \phi_L)$ domain if only the lightest chargino
mass is known.  Illustrations of such purely 
theoretical consistency constraints will be given
in the next sections.  

\subsection{Neutralino sector}

Let us now turn to the de-diagonalization of the neutralino sector. 
The question we want to answer analytically here, is how to 
determine in general $\vert M_1 \vert$ and $\Phi_{M_1}$ (as well as  
two remaining neutralino masses), for given  $M_2$, $\vert\mu\vert$, 
$\Phi_\mu$, $\tan  \beta$
and two arbitrary input neutralino masses $M_{N_1}$, $M_{N_2}$\footnote{In 
what follows
we denote the two input neutralino masses by $M_{N_1}$, $M_{N_2}$
for definiteness and assume them to be the lightest and next to lightest.
It should be clear, however,  that any two masses among the four
will can be equivalently used as input.}. At this stage $\mu$
and $M_2$ may be either obtained from the previous procedure, eqs. 
(\ref{mum2}--\ref{cphmu}), or known by any other mean. Clearly,
once the latter $M_1$ reconstruction algorithm is obtained, it can
easily be combined with the one of the previous section to determine 
$M_1, M_2$ and $\mu$ directly from input values of the physical parameters
$M_{\chi_1}$, $M_{\chi_2}$, $\phi_L$, $M_{N_1}$ and $M_{N_2}$, 
as will be illustrated in section 3.4.\\   
The neutralino mass matrix with the relevant phases reads
\beq
\label{Mneutralino}
M = \left(
  \begin{array}{cccc} |M_1| e^{i\Phi_{M_1}} 
& 0 & -m_Z s_W \cos\beta & m_Z s_W \sin\beta  \\
  0 & M_2 &  m_Z c_W \cos\beta & -m_Z c_W \sin\beta  \\
 -m_Z s_W \cos\beta & m_Z c_W \cos\beta & 0 & -|\mu| e^{i\Phi_\mu} \\
m_Z s_W \sin\beta & -m_Z c_W \sin\beta & -|\mu| e^{i\Phi_\mu} & 0 
\end{array} \right)
\eeq 
Since $M$ is now complex and symmetric but not hermitian, it cannot generally be
diagonalized through a similarity transformation (see eq.(\ref{nonsimili})).
In order to proceed as in ref\cite{inoinvreal}, 
we should rather consider the hermitian matrix
$M^\dagger M$,  and use the following four 
invariants under a similarity transformation: 
\begin{eqnarray}
\label{4inv}
& & T \equiv Tr M^\dagger M = \sum_{i=1,4} M^2_{N_i}\;;  \nn \\  \nn
& & \frac{(Tr M^\dagger M)^2}{2} - \frac{Tr ((M^\dagger M)^2)}{2} =
\sum_{i\neq j} M^2_{N_i} M^2_{N_j}\;; \\ \nn
& & \frac{(Tr M^\dagger M)^3}{6} - \frac{Tr M^\dagger M 
\,\,\, Tr ((M^\dagger M)^2)}{2} + \frac{Tr ((M^\dagger M)^3)}{3} =
\sum_{i\neq j \neq k} M^2_{N_i} M^2_{N_j} M^2_{N_k} \;;   \\ \nn
& & D \equiv Det M^\dagger M  = \defprod M^2_{N_i} \;;\\
\end{eqnarray}
where now the $M^2_{N_i}$ are the (squared) mass eigenvalues
 satisfying the eigenvalue equation
\beq
det(M^\dagger M -M^2_N I)
\eeq
with coefficients of  $(M^2_N)^i$ given by 
(\ref{4inv}) for $i=1,..4$. \\

After some lengthy but straightforward algebra, 
we obtain the explicit expressions for (\ref{4inv}) 
as a function of the various relevant parameters.
Following then a procedure similar to the one discussed in the appendix
of ref. \cite{inoinvreal}, but with the obvious replacements 
$M \to M^\dagger M$ and $M_{N_i} \to M^2_{N_i}$, 
we can solve this over constrained system analytically 
to give $|M_1|$, $\Phi_{M_1}$ and the two remaining 
physical neutralino masses. It is convenient
in a first stage to give the form of the equations controlling $Re[M_1]$, $Im[M_1]$:
\beqn
\label{M1sol}
a_1 Re[M_1]+ b_1 Im[M_1] +c_1 |M_1|^2 = d_1  \nn \\ 
a_2 Re[M_1]+ b_2 Im[M_1] +c_2 |M_1|^2 = d_2   \nn \\
\eeqn
where the coefficients $a_i$, $b_i$, $c_i$ and $d_i$ are functions of the 
parameters
$|\mu|$, $\Phi_\mu$, $M_2$, $M_{N_1}$, $M_{N_2}$ and $\tan\beta$. Once
$Re[M_1]$ and $Im[M_1]$ have been determined from (\ref{M1sol}), the
remaining unknown (squared) neutralino masses $M_{N_3}$, $M_{N_4}$
are obtained as the two solutions of a quadratic equation: 
\beq
\label{mn2mn3}
       M^2_{N_{3,4}} = \frac{1}{2}\left[ T -M^2_{N_1}-M^2_{N_2}
\mp \left( (T -M^2_{N_1}-M^2_{N_2})^2 -4 D/(M_{N_1}
M_{N_2})^2\right)^{1/2}\right]
\eeq   
where the trace $T$ and determinant $D$ are given as functions of
$\mu$, $M_1$, $M_2$ from eqs.(\ref{4inv}). 
 The algebraic expressions for $|M_1|$, $\Phi_{M_1}$ are not particularly
telling. We thus refrain from giving here more than their generic form
explicitly (their fortran encoding does not take, though,
 more than a couple of dozens of lines). \\
Rather, the more interesting characteristic features
of this neutralino inversion is that, quite similarly to the simpler
chargino case discussed above, non-trivial constraints among the 
physical masses arise, simply because the system (\ref{M1sol})
cannot always have a solution for any {\em arbitrary} 
$M_{N_1}$, $M_{N_2}$ input. More precisely, the system is (conditionally)
either over constrained and has no consistent solution; 
or has a twofold solution. 
This again simply reflects that not all possible $M_{N_1}$, $M_{N_2}$ values
can be consistently reached when taking into account that $|\cos\Phi_{M_1}|,
|\sin\Phi_{M_1}| \leq 1$. (For instance it is clear that the 
$\Phi_{M_1}$ phase freedom
cannot compensate for arbitrary values of $|\mu|$, $M_2$ or $|M_1|$, 
which more essentially determine the magnitude of the two neutralino
masses). 
In the specific form taken by the solution of the system 
(\ref{M1sol}) this issue is entirely driven by the 
positivity condition of the radicant $\Delta_N$, entering the quadratic
solution for $Re[M_1]$:

\begin{equation}
\Delta_N \geq 0 \label{radipos}
\end{equation}

\beq
Re[M_1] = \frac{\beta \pm \sqrt{\Delta_N}}{2\alpha}
\eeq

\beq
Im[M_1] = \gamma Re[M_1] + \delta 
\eeq

where $\Delta_N, \alpha, \beta, \gamma$ and $\delta$ have a complicated 
dependence on  $|\mu|$, $\Phi_\mu$, $M_2$, $M_{N_1}$, $M_{N_2}$ and
$\tan\beta$,  which we do not give here.
[The quadratic structure of the solutions is due to the 
special form of (\ref{M1sol}) where the non-linear terms appear only in 
$|M_1|^2$, otherwise one would have expected a quartic structure.]  
Accordingly, whenever the system is consistent, {\sl i.e.} eq.(\ref{radipos})
satisfied, there are two solutions. It may seem at first sight  
that this ambiguity occurs only as an artifact of  
``squaring" the mass matrix in our method. This is not so, 
for instance even if one considers both $M_1$ solutions afterwards in the 
actual (unsquared) mass matrix (\ref{Mneutralino}),
the ambiguity is not resolved but will be simply reflected in 
some phases of the mass eigenvalues, or equivalently
in  the couplings of the corresponding neutralino species.   
It follows that this twofold ambiguity on $M_1$, intrinsic to the 
case where only two
neutralino masses are input,  may in fact be
resolved by a study of the corresponding neutralino pair 
production cross-section which we consider in details
in section 4.\\

\subsection{Reconstruction of $M_1$, an illustration}
\begin{figure}[htb]
\vspace{-2cm}
\begin{minipage}[t]{19.cm}
\setlength{\unitlength}{1.in}
\begin{picture}(1.2,1)(0.8,10.3)
\centerline{\epsffile{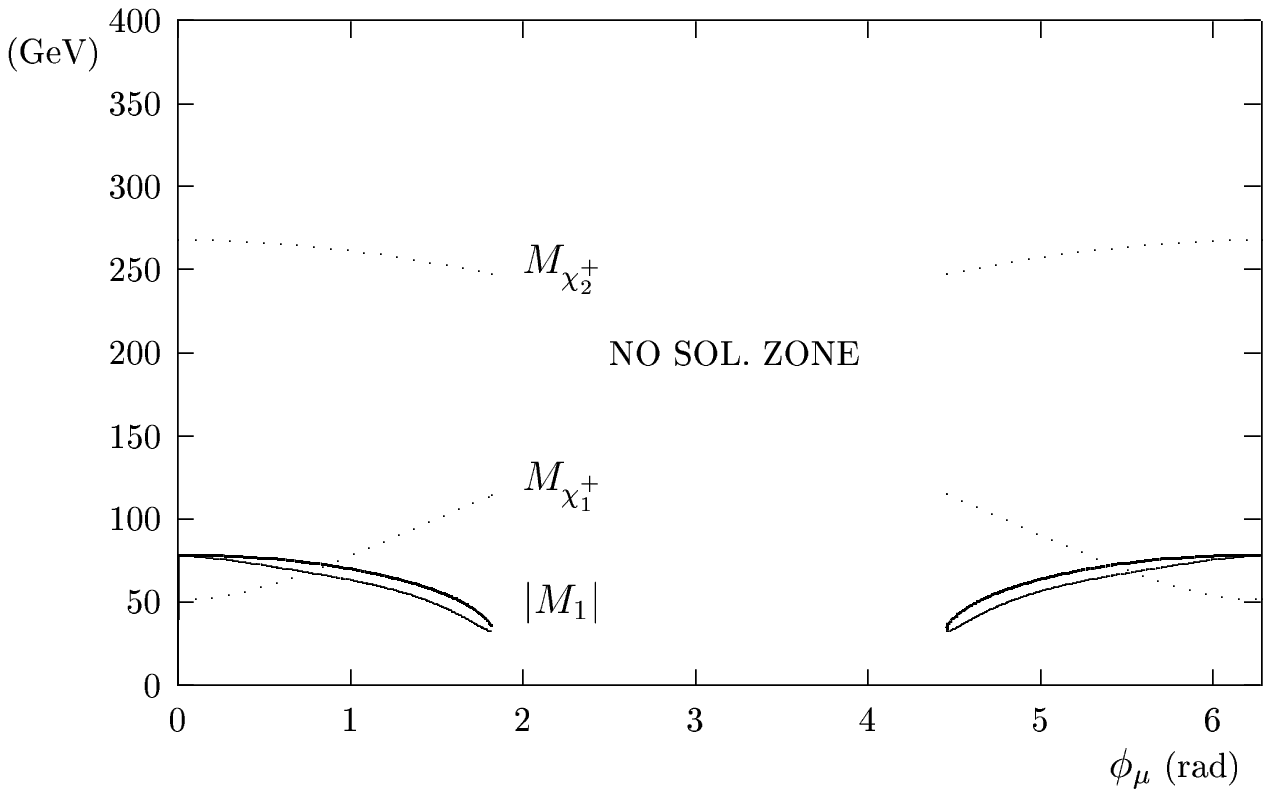}}
\end{picture}
\end{minipage}

\vskip 7cm

\begin{minipage}[b]{19.cm}
\setlength{\unitlength}{1.in}
\begin{picture}(1.2,1)(0.8,10.3)
\centerline{\epsffile{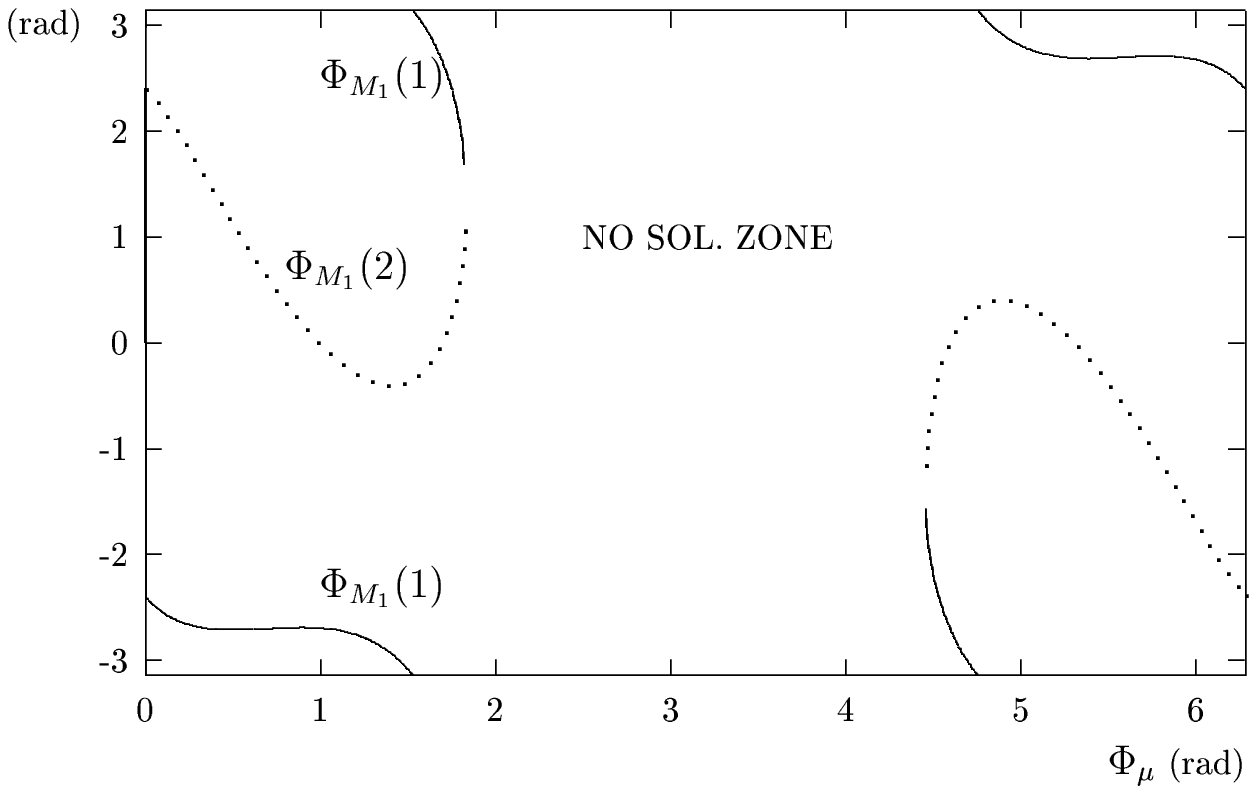}}

\end{picture}
\end{minipage}
\vspace{9cm}
\caption{\label{fig1} the twofold $|M_1|$ (top figure) and $\Phi_{M_1}$
(bottom figure) reconstruction with input choice $|\mu| =$ 100 GeV,
$M_2$ = 120 GeV,  $M_{N_1}=40$, $M_{N_2}=80$;  $\tan\beta=2$. Also
shown are the corresponding chargino mass values.} 
\end{figure}

To put more flesh on the previous algebraic construction,
we shall illustrate here some scenarios with typical input choices,  
and study the consequences on the relevant parameter reconstruction.
Let us first consider for simplicity the pure neutralino parameter
$|M_1|$, $\Phi_{M_1}$ reconstruction, thus taking  
$|\mu|$, $\Phi_\mu$ and $M_2$ as input. Even though this is not 
a reconstruction of $M_1$ directly from physical parameters, 
this partial algorithm has some interest before we consider a 
complete reconstruction from physical chargino and neutralino  masses.  \\
In fig. \ref{fig1} we chose typical input values $|\mu| =$ 100 GeV, 
$M_2 =$ 120 GeV (and $\tan\beta = 2$),
such that one of the charginos is relatively light,
$M_{\chi_1} \simeq $ 50--120 GeV, very roughly within the   
LEP2 reach or exclusion range~\footnote{One should, however, keep in mind that
the existing bounds\cite{PDG98,inobound,GDR-MSSM}, 
on top of being model dependent, have been 
derived assuming real parameters.}. 
We plot the resulting solutions for $M_1$ ($|M_1|$ in fig
1a (top) and $\Phi_{M_1}$ in fig. 1b (bottom)), as a function of the
phase $\Phi_\mu$, and for fixed $M_{N_1} =$ 40 GeV and $M_{N_2} =$ 80 GeV
input. (Note that in fig. 1b, the discontinuity in one of the $\Phi_{M_1}$
solution is not physical, being only an artifact of the definition of
the phase modulo $2\pi$. Our convention is such that 
$-\pi < \Phi_{M_1} < \pi$).  The shape of
the different plots is rather generic, and several characteristic
features can be noticed.  First, there is the rather important
central zone where no solution for $M_1$ exists, corresponding
 to the consistency
constraint (\ref{radipos}) not being fulfilled. 
For the input choice in fig. \ref{fig1},
this no-solution zone covers about one third of the possible $\Phi_\mu$
values, which is not so restrictive, but it should be noted that 
a relatively moderate change in the value of either $M_2/|\mu|$,
($\tan\beta$), or $M_{N_1}/M_{N_2}$ 
may easily reduce considerably the consistent region,
and quite often no consistent $M_1$ can be found for any 
$\Phi_\mu$ within a relatively 
important domain of the input parameter space. 
This will be illustrated more quantitatively below with ``scanning" plots.
One can already see the sensitivity
of the consistency relation (\ref{radipos}) to those changes
in fig 2a, 2b, where a larger $\tan\beta$ and a   
lower $M_{N_2}$ neutralino mass have been
chosen.

In the $\Phi_\mu$ 
range where $M_1$ solutions do exist, the plots in fig. 1a,b 
and 2a,b clearly exhibit the twofold
ambiguity on $M_1$ (except obviously at the $\Phi_\mu$ value(s)
such that $\Delta_N = 0$, the boarder of the no solution zone, where
$M_1$ is unique and the two distinct curves should meet.)
For the
input parameter choice in figs. 1 and 2, one observes that the two 
solutions are much more distinct for $\Phi_{M_1}$, than for $|M_1|$:
this is rather generic, although larger differences in the two $|M_1|$
solutions could occur, for instance for very different $M_{N_1}$,
$M_{N_2}$ values.  In fact, as is
clear from the figures,  the order of magnitude of $|M_1|$ is
essentially determined by the values of $|\mu|$, $M_2$ (or
$M_{\chi_1}$) $\simeq {\cal O}$ (100 GeV), while the $\Phi_\mu$ variation
here has a rather
moderate influence on $|M_1|$.  For example in fig. 1a, $|M_1|$ varies
as a function of $\Phi_\mu$ roughly between
a minimal ($\simeq M_{N_1}$) and a maximal ($\simeq M_{N_2}$) value, but
that property is somewhat accidentally due to the relatively close
values of $|\mu|$ and $M_2$ chosen here. When $M_2/|\mu|$
is larger (or smaller), we observed that the consistent values of $M_1$
are very close to the lightest $M_{N_1}$ input values, although  
it is hard to derive a simple generic behaviour.  

\subsection{Complete gaugino inversion, an illustration}
We will now illustrate the merging of the two separate algorithms 
described in previous sub-sections 3.1 and 3.2 for the chargino and
neutralino sectors. The outcome is thus the values of
$|\mu|$, $\Phi_\mu$,  $M_2$, $|M_1|$ and $\Phi_{M_1}$ as direct
expressions of the chargino masses and mixing angle $\phi_L$, and of
the two arbitrary neutralino masses (plus  $\tan\beta$). \\
In fig. 3a and 3b are plotted the output modulus and  
phases respectively, as functions of the physical 
chargino mixing angle $\phi_L$ for typical input chargino and neutralino 
masses (see figure captions), the lightest
being roughly close to present experimental
exclusion limits (although again, the exclusion analysis performed assuming
all parameters real may not apply here). 
The non-trivial correlations among the chargino and
neutralino masses appear now explicitly. First, and rather 
generically, the domain where 
both sectors can consistently exist is quite narrow as a function
of $\phi_L$ (the right boarder zones in 
fig. 3a,b are first eliminated from 
inconsistency with the relations      
(\ref{cphLbound}), while the condition $\Delta_N \ge 0$ eliminates most
of the remaining $\phi_L$ domain, leaving the relatively narrow
$M_1$ solution zones). Again, relatively moderate
changes in the input mass values may easily result in narrower
or even empty solution zones. For instance, changing 
$M_{\chi_1}$ only from 80 to 100 GeV, for the same values of the other
input parameters, gives no consistent solutions. Another example is
fig. 4 which illustrates the same reconstruction as in fig. 3 but for a
slightly higher $\tan\beta$ value. In summary, plots such
as the ones in fig. 3 and 4 show that the inversion provides in a direct way
the kind of correlations one is expecting among the neutralino masses,
once we know the chargino spectrum (or vice-versa). We could obviously
have illustrated the same kind of information by choosing for instance one of
the physical masses to vary, instead of $\phi_L$: the plots would be
different but would qualitatively reflect the same  strong
correlations. This will be illustrated more systematically in the next
sub-section.

\subsection{Reconstruction from partial input}
We consider now a less optimistic situation where only a partial knowledge of 
the physical input parameters is assumed, and illustrate the 
kind of information that can be
retrieved in this case. Accordingly in figs. 5--8 some input masses are
fixed while other and/or mixing angles are randomly 
scanned within a reasonable range for the masses 
(see figure captions). 
First, in fig.5 we fix the two chargino masses and $\tan \beta$,
and vary $\phi_L$ between  
$\phi_L \simeq 0.37$ and $\phi_L \simeq 0.51$ (rad), which corresponds 
to the first consistent solution zone in fig. 3a,b. 
Condition (\ref{radipos})  implies 
the pattern of correlation among the physical neutralino masses,
represented by the dotted regions (``butterflies") in the figure. 
More precisely, a given set of the four neutralino masses is consistent
only if any pair $(M_{N_i}, M_{N_j})$ of these masses corresponds to a
point lying on one of the ``butterflies". 
Definite consequences follow from this requirement such as the fact that each 
of the three allowed branches (along say the y-axis) can host only {\sl one}
pair $(M_{N_i}, M_{N_j})$. This allows to make qualitative statements about
the neutralino spectrum relative to that of the charginos. 
For instance, with the specific input chosen in fig.5, close inspection
shows that there should
always be one (and only one) neutralino heavier than the heaviest chargino,  
and that $M^+_{\chi_1} - M_{N_1}$ gets smaller when 
$M_{N_4} -M^+_{\chi_2}$ gets larger, and vice-versa 
($M_{N_1}$ is the LSP and $M_{N_4}$ the heaviest neutralino in our
conventions).  If $\tan\beta$ increases, the
dotted  butterflies are simply moving up or down along the diagonal 
$M_{N_i}=M_{N_j}$ line, as illustrated for the choice of parameters in
fig. 6: thus the (anti)correlation property between the lightest and
heaviest mass splitting remains, but for larger $\tan\beta$ values there
is room for the four neutralinos being lighter than $M^+_{\chi_2}$.
Finally, note that for the input parameter choice of fig. 5 and 6, 
there is no consistent solution when $\tan\beta \gsim 8$. \\  
We stress
again that these correlations in fig. 5 and 6  are only due to the
theoretical  consistency relation (\ref{radipos}), and are simply obtained
from scanning over the two input neutralino mass ranges,
calculating (\ref{radipos}) only once for each $M_{N_i}$, $M_{N_j}$,
(and eventually $\phi_L$) input.  On top of
this, one could easily add any direct experimental constraints
on the LSP and other  gaugino masses \cite{PDG98,inobound,GDR-MSSM} etc
(which we however refraine from investigating here, since it is difficult
at present to  infer
which experimental mass limits 
remain valid when non-zero phases are assumed). 
From a practical point of view, it may be useful to compare 
how figs. 5 and 6
could be obtained from a more conventional procedure. Accordingly, one
would have to scan over the five-parameter space 
$M_2$, $|\mu|$, $\Phi_\mu$, $|M_1|$, $\Phi_{M_1}$, with $\tan\beta =2$
($\tan\beta =5$),
diagonalizing the complete chargino and neutralino matrices for each input
point, and finally selecting only those output corresponding to 
the required physical chargino and neutralino masses. Even if our method
is not intended to replace conventional procedures of e.g. fitting to
the  data, it
can be in many practical situations 
an appreciable ``theoretical consistency" guide with a substantial gain in
simulation time. \\

Next, we consider another application of the gaugino
spectrum algebraic inversion, where we try to illustrate the kind of
minimal data needed to e.g. draw definite conclusions
about non-zero phases. Figs. 7 and 8 give typical ``contour" plots in
the ($|\mu|, \Phi_\mu$) slice of the basic parameter space, when
assuming that only the lightest chargino and the two neutralino masses
are fixed, while the heavy chargino mass is basically unknown, and
also $\tan\beta$ varies within some range. Figs. 7 and 8 illustrate two
different choices for the chargino and the two neutralino masses in
this scenario (see figure caption). For the input chosen here, one can
see that the ``data" definitely imply large non-zero $\Phi_\mu$, and
this statement does not depend on the specific value of $M_{\chi_1}$, as
shown by  comparing fig. 7 and 8 input. However, when increasing
$\tan\beta$ or the splitting between the two neutralino masses, more and more 
points appear with $\Phi_\mu$ close to zero.  Qualitatively,
one can summarize this behaviour by noting that
there exists a well-defined  correlated region in the chargino-neutralino 
mass parameters, such that for small enough $\tan\beta$, $\Phi_\mu$ is
necessarily large (the precise value of $\tan\beta$
depending on the precise values of the neutralino mass splitting). It
should be thus interesting to simulate this scenario in a more
systematic way, taking
into account also independent constraints on the phases 
$\Phi_\mu$, $\Phi_{M_1}$~\cite{edme}--\cite{Savoyetal}, on  
$\tan\beta$, etc, which is however beyond
the scope of the present paper.

\section{The $e^+e^- \to \chi^0_i\chi^0_j$ cross-section with phases} 
In this section, we recalculate and illustrate
 the neutralino pair production in the complex parameter case, in
$e^+e^-$ collisions. Our  motivation is twofold:\\ 

i) once non-zero $M_1$, $\mu$ phases are assumed
in the gaugino sector, all predictions for chargino, neutralino
production processes will be changed, and it is crucial to analyze the
relevant production cross-section   
with the general phase dependence. This was done 
for chargino pair production 
in ref\cite{choi1,choi2} precisely to extract the chargino  
sector complex parameters. As for the neutralino pair production
in $e^+e^-$ collisions, the
cross-section with non-trivial phases was calculated 
in the past~\cite{Petcov} and
CP-violation effects in the neutralino pair
production and decay was also considered in some
details~\cite{een1n2phase,n1n2lc}. 

ii) As was stated previously, the
reconstruction algorithm, using only two neutralino mass input ,
leads to a twofold ambiguity in the resulting $M_1$ values.
We shall illustrate, however, that the above-mentioned
$M_1$ ambiguity can be in most cases resolved from the
corresponding neutralino 1,2 production total cross-section,  if the
latter is measured with a reasonable accuracy. As a by-product of the
inversion procedure, we can also obtain direct correlations between 
the chargino and neutralino pair production cross-sections.

\subsection{Analytical expressions} 
\setcounter{equation}{0}
The neutralino $i,j$ 
pair production cross-section in $e^+e^-$ collision at tree-level
proceeds through s channel $Z, \gamma$ and t(u)-channel sneutrino
$\tilde{\nu}_e$ exchange. 
The cross-section for the real
parameter case was calculated in ref.\cite{bartl1,ambrosanio}, and for
the complex
parameter case in refs~\cite{een1n2phase}. Here we have recalculated
this process independently for the full phase dependence in our
conventions. In terms of the matrix elements $Z_{ij}$ diagonalizing the
neutralino mass matrix (\ref{Mneutralino}): \beq Z^* M Z^{-1}
\equiv M_{diag}\;, \label{nonsimili}\eeq
which are given functions
of the parameters $\mu$, $M_1$, $M_2$, $\tan\beta$,
the cross-section reads:
\beq
\sigma(e^+e^- \to \chi^0_i\chi^0_j) = \frac{(2 - \delta_{i j})}{2} (\sigma_Z  + 
\sigma_{\tilde{e}} + \sigma_{Z \tilde{e}} ) 
\eeq 
where the three contributions to the integrated cross-section are given by

\begin{equation}
\label{sigz}
\sigma_Z= \frac{q }{8 \pi s \sqrt{s}} \frac{g^4 (L_e^2 + R_e^2)}
{\cos^4_W} 
               |D_Z (s)|^2 
              \{  I_1
                \; |O_{j i}^{''L}|^2  
            \;\; -2 s \;m_i m_j\; \Re [(O_{j i}^{''L})^2] \}
\end{equation}

\begin{equation}
\label{sige}
\sigma_{\tilde{e}} =   \frac{q }{32 \pi s \sqrt{s}} g^4 
             \{ |f_{l_i}^L|^2 |f_{l_j}^L|^2  I_3(m_{\tilde{e}_L})
         - s \;m_i m_j\; \Re [(f_{l_i}^L)^2 ({f_{l_j}^{L}}^*)^2]
                       \;\;   I_5(m_{\tilde{e}_L})
                          + ( L \to R) \}
\end{equation}

\begin{eqnarray}
\label{sigze}
\sigma_{Z \tilde{e}} = \frac{q }{8 \pi s \sqrt{s}} \frac{g^4}
{\cos^2_W} 
           \Re[D_Z (s)]
       \{ L_e \;\; (\Re[{f_{l_j}^{L}}^* f_{l_i}^{L} O_{j i}^{''L}]
                   \; I_2(m_{\tilde{e}_L})
             - s \;m_i m_j\; \Re[{f_{l_j}^{L}}^* f_{l_i}^{L} {O_{j i}^{''L}}^*] 
I_4(m_{\tilde{e}_L}) ) &&\nonumber \\ 
    \;\;\; \;\;\;\  \;\;\;\; \;\;\;\; -R_e \;\;( \Re[{f_{l_j}^{R }}^* f_{l_i}^{R } {O_{j i}^{''L}}^*] \; I_2(m_{\tilde{e}_R })
                      - s \;m_i m_j\; \Re[{f_{l_j}^{R }}^* f_{l_i}^{R } O_{j i}^{''L}]
I_4(m_{\tilde{e}_R }) )\}&& \nonumber \\ 
\end{eqnarray}

where $m_i$, $m_j$ designate the two neutralino {\sl physical} masses,
and the various couplings are defined as 
\beqn
\label{fLR}
& & f_{l_i}^{L} = \sqrt{2} \:[ (T_{3l} -e_l)\: \tan\theta_W\: Z_{i1}
- T_{3l} \; Z_{i2} ]\nn \\ 
& & f_{l_i}^{R} = \sqrt{2} \: e_l \tan\theta_W \: Z^*_{i1} \nn \\
\eeqn

\beq
L_l = T_{3l} -e_l\sin^2\theta_W\;,\;\;\;\;R_l =-e_l\sin^2\theta_W,
\eeq
and
\beq
O_{i j}^{''L} = -\frac{1}{2} Z_{i 3} Z^*_{j 3} + 
             \frac{1}{2} Z_{i 4} Z^*_{j 4} =  {O_{j i}^{''L}}^*
\eeq
is the coupling of the $Z$ to neutralino $i,j$ 
with $ O_{i j}^{''R} = - {O_{i j}^{''L}}^*$ 
(see the second ref. in \cite{R1} for more details, but note a typo in eq.C77 
therein.)

The kinematical variables are 
\begin{eqnarray}
&&E_i = \sqrt{q^2 + m_i^2} \nonumber \\
&&q^2= \frac{1}{4 s} ( s - ( m_i + m_j)^2 ) ( s - ( m_i - m_j)^2) \nonumber \\
&&D_Z(s)= \frac{1}{s - m_Z^2 + i m_Z \Gamma_Z} \nonumber \\
\end{eqnarray}
and the expressions for the integrals $I_n$ appearing in eqs.
(\ref{sigz}--\ref{sigze}) are collected in the appendix, where we also
display the relevant expressions for the amplitudes. 
Note that the complex phase factors 
in the mass eigenvalues
(usually denoted by $\eta_i$~\cite{Petcov}--\cite{bartl1}) 
are in our procedure automatically taken into
account in eqs. (\ref{sigz}--\ref{sigze}), in both CP-conserving
and CP-violating cases, through the redefinition
\beq
Z_{jk} \to Z_{jk} e^{i \theta_j/2}
\eeq   
when the mass eigenvalue of the neutralino $j^{th}$ species picks up a $\theta_j$ phase.
 
Our expressions agree analytically with ref. \cite{bartl1} 
in the real case 
and with ref. \cite{Petcov} in the complex case. 
We obtained as well very good numerical
agreement with the values of the cross-section displayed in ref.
\cite{ambrosanio}.

\subsection{Illustrations of $\chi^0_1\chi^0_2$ production
cross-section} In fig. \ref{nino_fig1} we plot the $e^+e^- \to 
\chi^0_1 \chi^0_2$ total cross-section
for the same choice of parameters corresponding to fig.
1a,b, and  two different choices of the selectron masses (see figure caption), 
for
a LEP2 energy of 190 GeV (top figure) and a future 
linear collider energy of 500 GeV. (For simplicity, we assume a negligible
mixing in the selectron sector, so that $m_{\tilde e_L}$ and 
 $m_{\tilde e_R}$ are the physical masses). Although our input 
parameter choice
is particular, the behaviour of the cross-section  is rather generic. 
As a first general observation, this plot illustrates the 
important sensitivity
of the total cross-section to the variation of $\Phi_\mu$ and $\Phi_{M_1}$
(the plots in fig. 9 are functions of $\Phi_\mu$ but $\Phi_{M_1}$
varies also for this input choice, according to our inversion algorithm,
as is clear from the corresponding fig. 1b.) For instance, for both 
choices of selectron masses illustrated in the figures, the values of the 
cross-section vary roughly by about 30-40 \%  
when the phases are varied within
their maximal possible range, both at LEP2 and linear collider energies. 
This is not too surprising, since although the two neutralino masses
(and the two selectron masses) are fixed, there is an important
sensitivity to the phases through the $Z$-neutralinos
and $e$-$\tilde e$-neutralino couplings entering the
cross-section formula, eqs. (\ref{sigz}--\ref{sigze}).
Now a rather unpleasant feature of the presence of non-zero phases
is that, depending on the selectron masses, one may have several different
$\Phi_\mu$ ($\Phi_{M_1}$) values giving 
the same total cross-section value, even though 
the two neutralino masses
and the selectron masses are fixed (see for instance 
the dotted lines in fig. 9, corresponding to the second $M_1$
solution at 190 GeV, for $m_{\tilde e_L}, m_{\tilde e_R} =$ 100,
120 GeV). These ambiguities on the phases may however be easily
resolved when looking at the corresponding values of the chargino
masses (provided the latter are known), whose corresponding  variation
with $\Phi_\mu$ is illustrated in fig. 1a. This clearly  stresses again 
the importance of looking at possible correlations between the neutralino
and chargino sectors in general.   
Note also that the cross-section
values in fig. 9 
are symmetric  with respect to $\Phi_\mu \to 2\pi -\Phi_\mu$ (which also
implies $\Phi_{M_1} \to -\Phi_{M_1}$, see e.g. 
fig. 1b).
Concerning now the $M_1$ reconstruction ambiguity, which is 
more specific to our inversion algorithm,  
the plots in fig. 9 also illustrate that the total
cross-section is generally quite sensitive to the twofold $M_1$ solution:
even though the two $|M_1|$ solutions of fig. 1a 
are very close to each others, the two solutions for the phases $\Phi_{M_1}$
are not, thus leading to the above mentioned sensitivity.
Therefore, provided the cross-section is sufficiently large to be
measured with a reasonable accuracy~\cite{LEPYR96,lcrep},
one should be able to resolve this ambiguity in our 
$M_1$ reconstruction procedure
rather easily, provided of course that
the selectron masses are also known. \\      

In fig. \ref{nino_fig2} we plot the $e^+e^- \to \chi^0_1 \chi^0_2$ 
total cross-section at LEP2 (top figure)
and future linear collider energies (bottom figure), 
but now as a direct function of the
chargino, neutralino masses and chargino mixing angle,  
for the same choice of parameters corresponding to fig. 3a,b and 
one choice of the selectron masses (see figure caption). The regions
with no plots in fig 10 simply correspond to the excluded $\phi_L$
range from consistency of the inversion, as discussed previously for the
 plots in figs. 3a,b.  
Again, the different plots illustrate the generically rather important
sensitivity of the cross-section to the phases (the plots in fig. 10
are functions of the mixing angle $\phi_L$, which corresponds
for fixed chargino and neutralino masses to varying $\Phi_\mu$ 
and $\Phi_{M_1}$, as is clear from the
corresponding fig. 3b). 
One should keep in mind that since the mixing angle $\phi_L$
(more precisely $\cos 2 \phi_L$)
can be in principle determined from the measurement of the chargino pair 
producion cross-section, plots of the neutralino pair production like in 
fig.10 may be viewed as a function of the chargino pair production, for fixed 
chargino and neutralino mass values. This illustrates further the strong 
correlations between the chargino and neutralino physical parameters. 
As for the ambiguities in our inversion algorithm, they come in
this case too from the twofold solution in the reconstruction of  $M_1$.
Again, given the generically quite different
cross-section values corresponding to the two $M_1$ solutions in fig. 10,
this ambiguity should be easily resolved through the measurement 
of this observable.

\section{Conclusion} 

In the present paper, we have derived a 
purely algebraic algorithm to reconstruct the (unconstrained) gaugino sector
Lagrangian parameters $\mu \equiv |\mu| e^{i\Phi_\mu}$, 
$M_2\equiv |M_2|$ and 
$M_1 \equiv |M_1| e^{i\Phi_{M_1}}$, directly from the physical
chargino and (some of the) neutralino masses.  
Our construction exhibits in a more direct and systematic way the
non trivial correlations among the physical chargino and neutralino
physical parameters, which exist even when the
maximal possible phase freedom of the unconstrained MSSM parameter space
is considered, and which may be very hidden or cumbersome to extract
in the more standard approach of 
systematic scanning over the basic parameters. Our approach 
should be useful in particular in 
the case where only a subset of the minimal required input is
available, and we illustrated with several typical such scenarios
the kind of theoretical consistency constraints that could result, in
addition to independent experimental constraints.\\

We have also recalculated the neutralino pair production
cross-section $e^+e^- \to \chi^0_i \chi^0_j$ within our framework
and phase conventions, and illustrate its sensitivity
to the phases of $\mu$ and $M_1$, which should be 
in particular useful
to resolve an ambiguity in the $M_1$ parameter reconstruction, intrinsic
to our algorithm. A careful comparison of the chargino and neutralino 
pair production together with the use of the inversion algorithm should
also give non trivial correlations among the different
chargino/neutralino masses and cross-section observables. Those
possible constraints  may be studied in a more systematic way, taking
into account also the additional direct or indirect constraints from
LEP data, measurements of the electric dipole moment, etc...\\ 

Finally, a refinement of our tree-level
algebraic inversion procedure would be 
to include radiative corrections. However, these
corrections are expected
to be generically much smaller for the gaugino masses~\cite{rcmass} than
for the chargino pair production cross-section~\cite{rcsig}, and as such,
would probably affect more the reconstruction algorithm 
of the chargino parameters than that of the neutralino ones.

{\large \bf Acknowledgments}\\
We are thankful to  S.Y. Choi, Fran\c cois Richard and Carlos Savoy
for useful discussions. 

\section*{Appendix: Neutralino pair production formulas}
\setcounter{equation}{0}
We collect here some formulas needed for the expression of the
neutralino $i,j$ pair production cross-section. The integrals 
in (\ref{sigz}--\ref{sigze}) read
\begin{eqnarray}
I_1 &= & \frac{2 s}{3} (q^2 + 3 E_i E_j) \nonumber \\
I_2[m] &= & \frac{1}{4}[ 
\frac{ ( (m_i^2 + m_j^2 - 2 m^2)^2 - s (E_i - E_j)^2)}{q \sqrt{s}} L[m]
-4 (-2 m^2 + \sqrt{s} (E_i + E_j) + m_i^2 + m_j^2) ] \nonumber \\
I_3[m] &= &  \frac{1}{2} [\frac{
8 (2 s (-q^2 +  E_i  E_j) - 
     \sqrt{s} ( E_i +  E_j) (-2 m^2 + m_i^2 + m_j^2) + 
     (-2 m^2 + m_i^2 + m_j^2)^2)}
{-4 q^2 s + (-2 m^2 - \sqrt{s} (E_i + E_j) + m_i^2 + m_j^2)^2 } \nonumber \\ 
      & +& \frac{2 L[m] (2 m^2 - m_i^2 - m_j^2)}{q \sqrt{s}} ]
\nonumber \\
I_4[m] &= & \frac{L[m]}{q \sqrt{s}} \nonumber \\
I_5[m] &=& \frac{-2 L[m]}{q \sqrt{s} (2 m^2 + \sqrt{s} (E_i +E_j) 
- m_i^2 - m_j^2)}\nonumber \\
L[m]&=& \log{\frac{- 2 m^2 + m_i^2 + m_j^2 - \sqrt{s} (E_i + E_j - 2 q)}{
              - 2 m^2 + m_i^2 + m_j^2 - \sqrt{s} (E_i + E_j + 2 q)} } 
\end{eqnarray}

The square amplitudes  for the neutralino $i, j$ pair production involve
$s-$, $t-$ and $u-$ channel exchanges. They read: 

\begin{eqnarray}
|T_s|^2  =  4 \frac{g^4}{c_w^4} |D_Z (s)|^2 (L_e^2 + R_e^2)
              \{ ( \xi_{i j }(t) + \xi_{i j }(u) )
                \; |O_{j i}^{''L}|^2  
            \;\; -2 s \;m_i m_j\; \Re [(O_{j i}^{''L})^2] \}&&
\end{eqnarray}

\begin{eqnarray}
|T_t + T_u|^2 & = & g^4 \{ |f_{l_i}^L|^2 |f_{l_j}^L|^2 (
                         {D_{\tilde{e}_L} (t)}^2 \xi_{i j }(t) +
                         {D_{\tilde{e}_L} (u)}^2 \xi_{i j }(u) ) \nonumber \\
       & &  -2 s \;m_i m_j\; D_{\tilde{e}_L} (t) D_{\tilde{e}_L} (u)
                          \Re [(f_{l_i}^L)^2 ({f_{l_j}^{L}}^*)^2]
                          + ( L \to R) \}
\end{eqnarray}

\begin{eqnarray}
&& 2 \Re[ T_s ( T_t + T_u)^*] = 4 \frac{g^4}{c_w^2}\; \{ \nonumber \\
&&     \;\;\;        L_e \; \{\;D_{\tilde{e}_L} (t)[\;\;
                     \xi_{i j }(t) \Re[ D_Z(s) f_{l_j}^{L} {f_{l_i}^{L}}^*
                               {O_{j i}^{''L}}^*] 
              -   s \;m_i m_j\; \Re[ D_Z(s) f_{l_j}^{L} {f_{l_i}^{L}}^*
                               O_{j i}^{''L}] \;\;] \; + \nonumber \\
&&\;\;\;\;\;\;\;\;\;\;  D_{\tilde{e}_L} (u)[\;\;
                     \xi_{i j }(u) \Re[ D_Z(s) {f_{l_j}^{L}}^* f_{l_i}^{L}
                               O_{j i}^{''L}] 
              -   s \;m_i m_j\; \Re[ D_Z(s) {f_{l_j}^{L}}^* f_{l_i}^{L}
                               {O_{j i}^{''L}}^*] \;\;]   \;\;    \} 
                      \nonumber \\
&& -R_e\; \{\; D_{\tilde{e}_R} (t)[\;\;
                     \xi_{i j }(t) \Re[ D_Z(s) f_{l_j}^{R} {f_{l_i}^{R}}^*
                               O_{j i}^{''L}] 
              -   s \;m_i m_j\; \Re[ D_Z(s) f_{l_j}^{R} {f_{l_i}^{R}}^*
                               {O_{j i}^{''L}}^*] \;\;] \; + \nonumber \\
&&\;\;\;\;\;\;\;\;\;\;  D_{\tilde{e}_R} (u)[\;\;
                     \xi_{i j }(u) \Re[ D_Z(s) {f_{l_j}^{R}}^* f_{l_i}^{R}
                               {O_{j i}^{''L}}^*] 
              -   s \;m_i m_j\; \Re[ D_Z(s) {f_{l_j}^{R}}^* f_{l_i}^{R}
                               O_{j i}^{''L}] \;\;]  \;\; \} \;\;\}  \nonumber \\  
\end{eqnarray}                               
where
\beq
\xi_{i j} (x) \equiv (m_i^2 - x )(m_j^2 - x)
\eeq

\beq
D_{\tilde{e}_{L, R}}(x) = \frac{1}{x - m_{\tilde{e}_{L, R}}^2 } 
\eeq
and the other functions and couplings as defined in section 4. The differential
cross-section is given by

\begin{equation}
\frac{d\sigma}{d \cos \theta}= \frac{q}{64 \pi s \sqrt{s}} ( |T_s|^2 +  2 \Re[ T_s ( T_t + T_u)^*] + |T_t + T_u|^2)
\end{equation}

\newpage

\newpage

\begin{figure}[htb]
\vspace{-2cm}
\begin{minipage}[t]{19.cm}
\setlength{\unitlength}{1.in}
\begin{picture}(1.2,1)(0.8,10.3)
\centerline{\epsffile{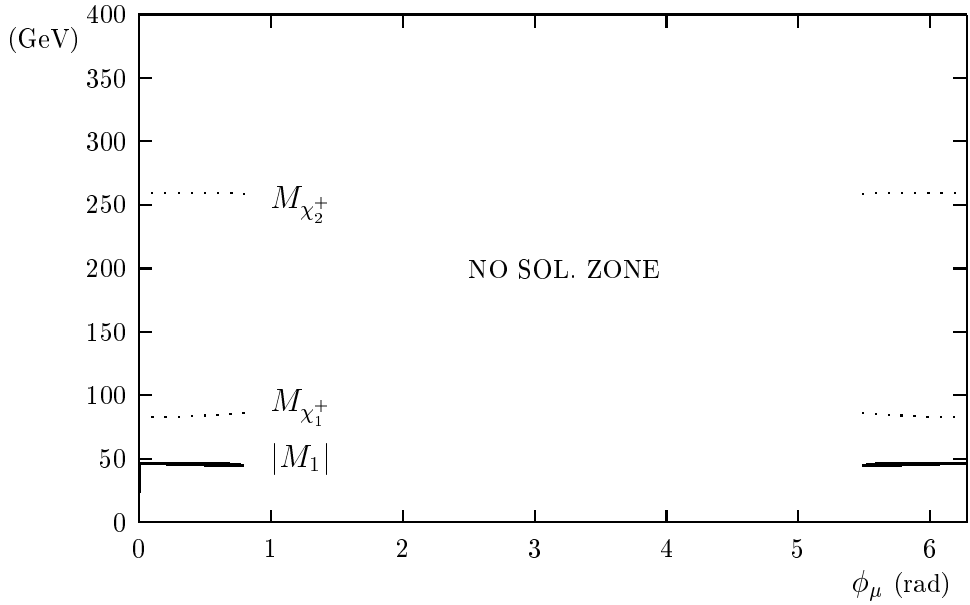}}
\end{picture}
\end{minipage}

\vskip 7cm

\begin{minipage}[b]{19.cm}
\setlength{\unitlength}{1.in}
\begin{picture}(1.2,1)(0.8,10.3)
\centerline{\epsffile{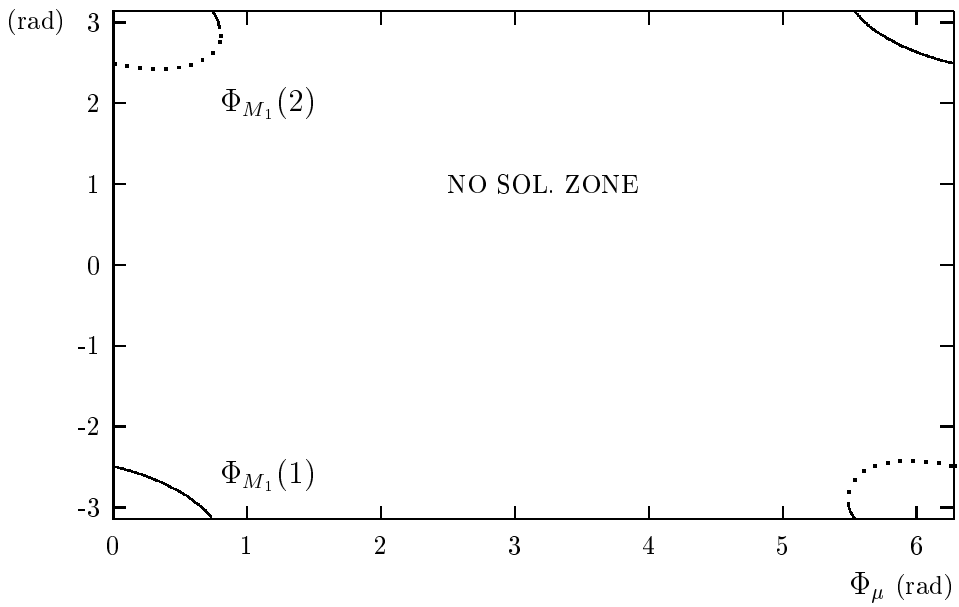}}
\end{picture}
\end{minipage}
\vspace{9cm}
\caption{ \label{fig2} same as fig. 1, with input choice $|\mu| =$ 100
GeV, $M_2$ = 120 GeV,  $M_{N_1}=40$, $M_{N_2}=60$;  $\tan\beta=10$.}
\end{figure}

\newpage

\begin{figure}[htb]
\vspace{-2cm}
\begin{minipage}[t]{19.cm}
\setlength{\unitlength}{1.in}
\begin{picture}(1.2,1)(0.8,10.3)
\centerline{\epsffile{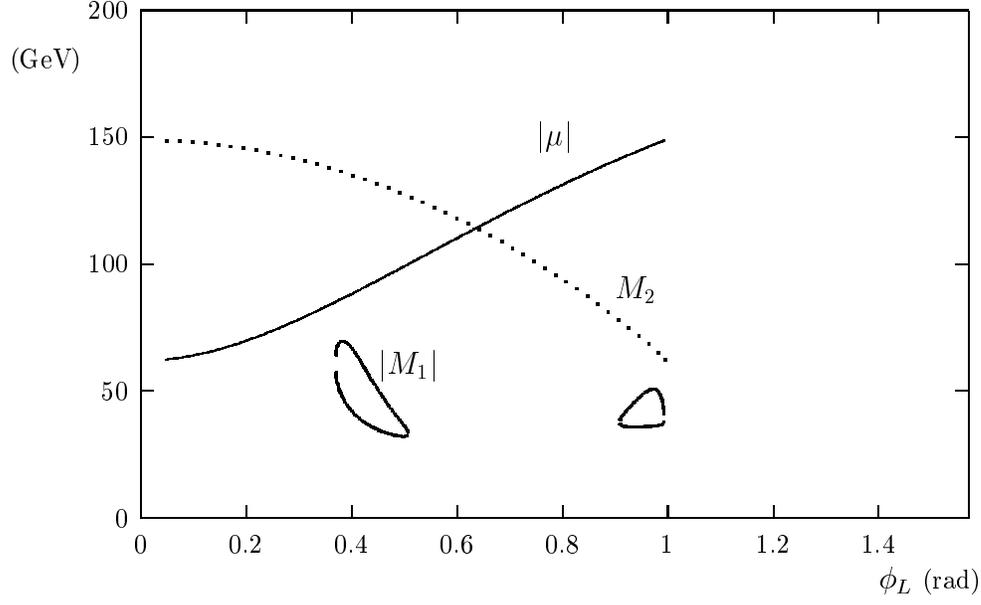}}
\end{picture}
\end{minipage}

\vskip 7cm

\begin{minipage}[b]{19.cm}
\setlength{\unitlength}{1.in}
\begin{picture}(1.2,1)(0.8,10.3)
\centerline{\epsffile{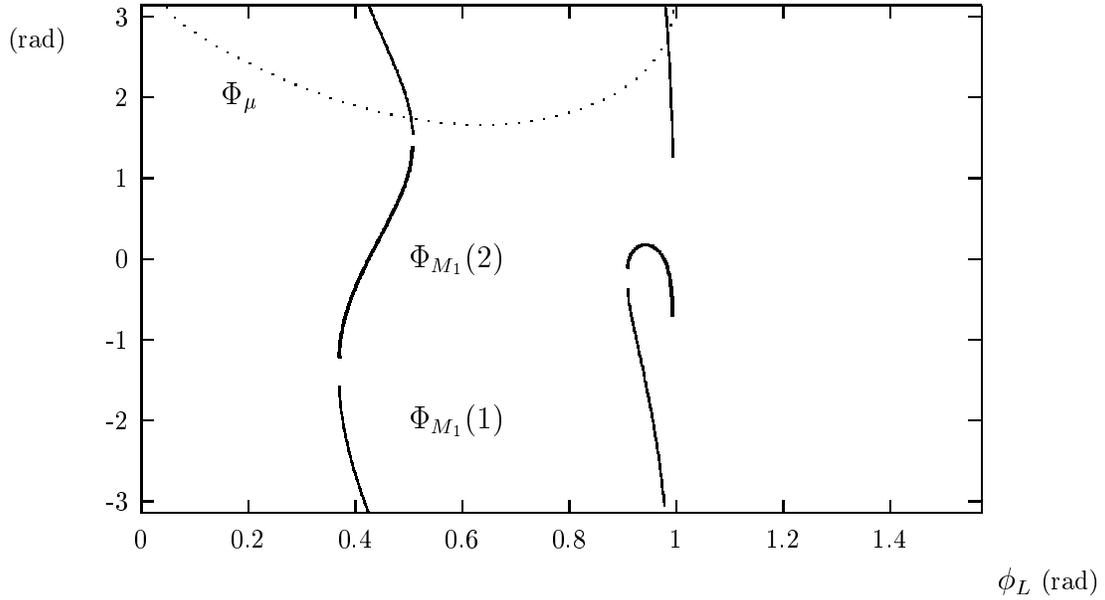}}
\end{picture}
\end{minipage}
\vspace{9cm}
\caption{ \label{fig3} $|M_1|$, $M_2$, $|\mu|$ (top figure) and
$\Phi_\mu$, $\Phi_{M_1}$ (bottom figure) full reconstruction with
physical input choice $M_{\chi^+_1} =$ 80 GeV, $M_{\chi^+_2} =$ 180
GeV,  $M_{N_1}=40$, $M_{N_2}=80$;  $\tan\beta=2$.} 
\end{figure}

\newpage

\begin{figure}[htb]
\vspace{-2cm}
\begin{minipage}[t]{19.cm}
\setlength{\unitlength}{1.in}
\begin{picture}(1.2,1)(0.8,10.3)
\centerline{\epsffile{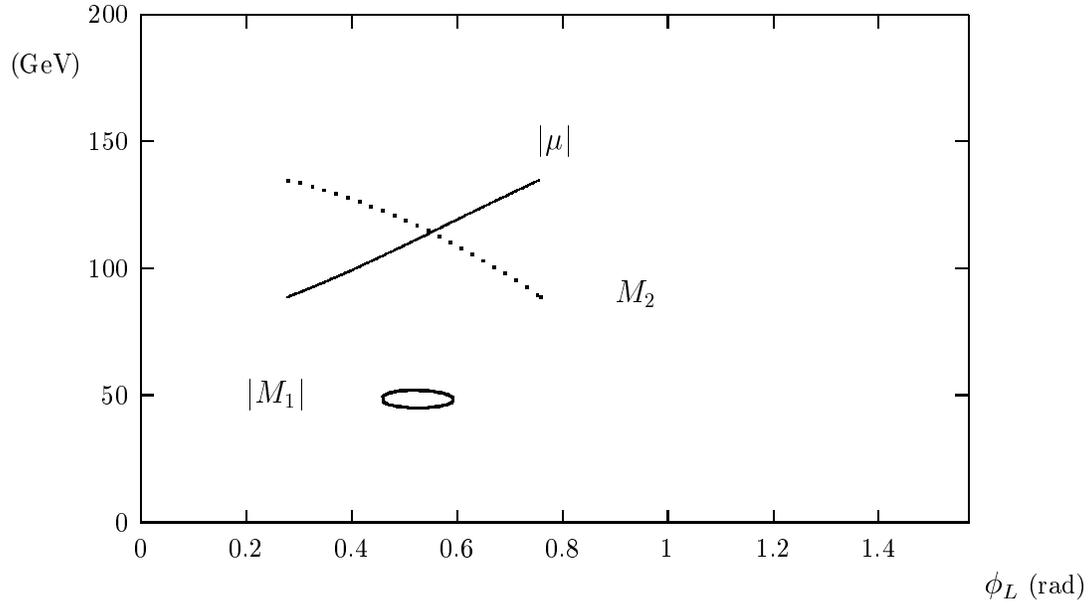}}
\end{picture}
\end{minipage}

\vskip 7cm

\begin{minipage}[b]{19.cm}
\setlength{\unitlength}{1.in}
\begin{picture}(1.2,1)(0.8,10.3)
\centerline{\epsffile{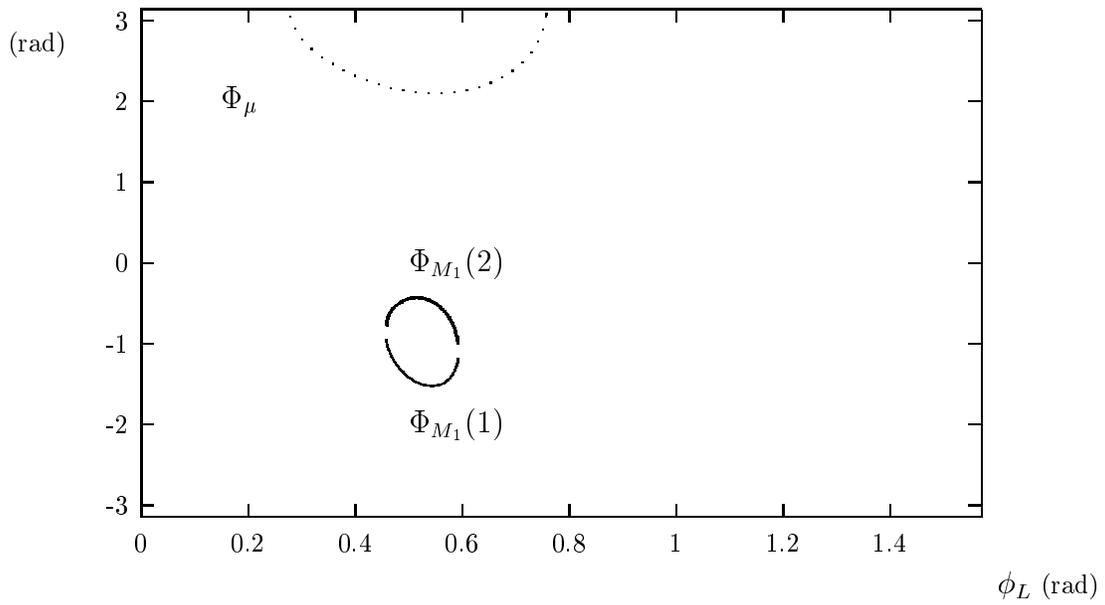}}
\end{picture}
\end{minipage}
\vspace{9cm}
\caption{ \label{fig4} same as fig. 3 with input choice $M_{\chi^+_1} =$
80 GeV, $M_{\chi^+_2} =$ 180 GeV,  $M_{N_1}=40$, $M_{N_2}=80$; 
$\tan\beta=5$.} 
\end{figure}

\newpage
\begin{figure}[htb]
\centerline{\epsffile{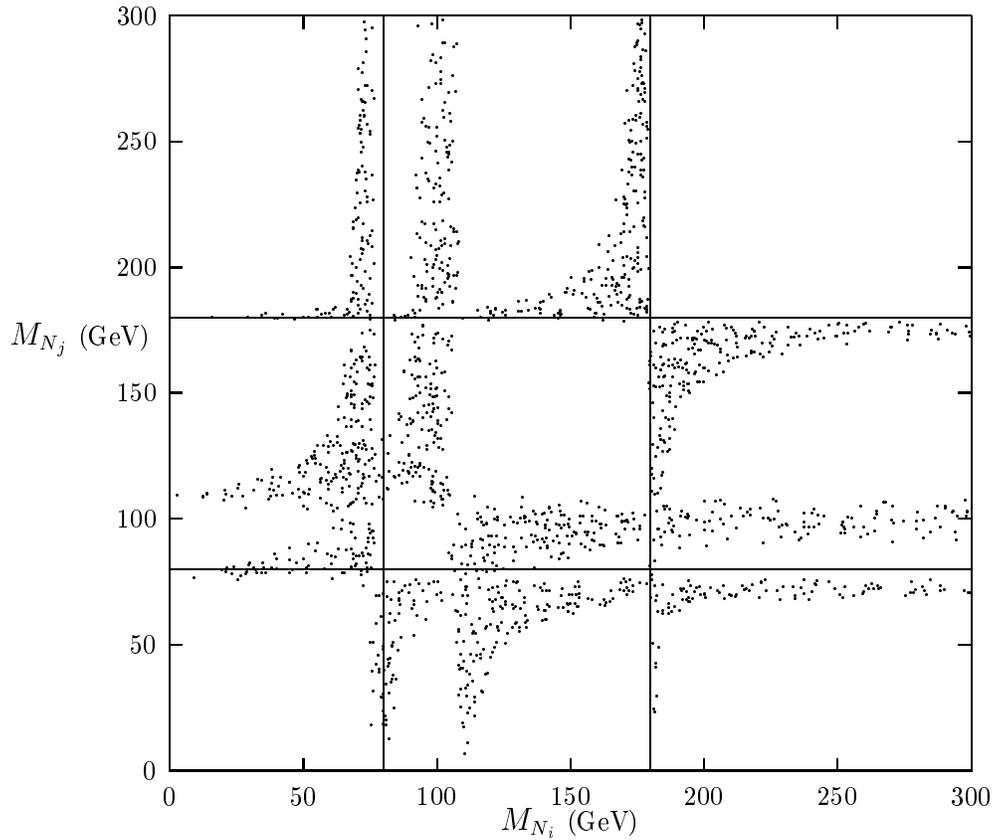}}
\vspace{-11cm}
\caption{ \label{fig5-1} Correlations between arbitrary
neutralino masses from consistency of the inversion. Input parameters: 
$M_{\chi^+_1} =$ 80 GeV, $M_{\chi^+_2} =$ 180
GeV (indicated by the two straight lines),  0.37 $< \phi_L(rad) <$ 0.52;  $\tan\beta= $2.}
\end{figure}

\newpage
\begin{figure}[htb]
\centerline{\epsffile{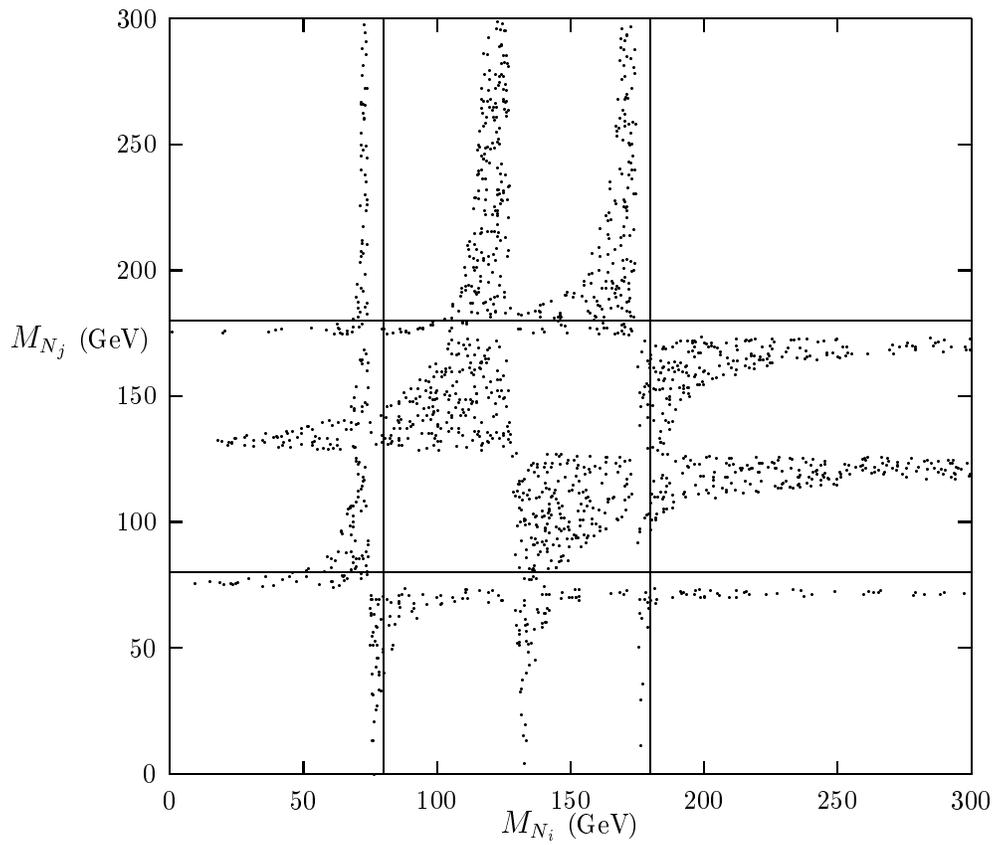}}
\vspace{-11cm}
\caption{ \label{fig5-2} same as fig. 5 for a different
input choice: $M_{\chi^+_1} =$ 80 GeV, $M_{\chi^+_2} =$ 180 GeV,  $
\phi_L(rad) =0.5 $ ;  $\tan\beta = 5 $} 
\end{figure}

\newpage
\begin{figure}[htb]
\vspace{-3cm}
\begin{minipage}[t]{19.cm}
\setlength{\unitlength}{1.in}
\begin{picture}(1.2,1)(0.8,10.3)
\centerline{\epsffile{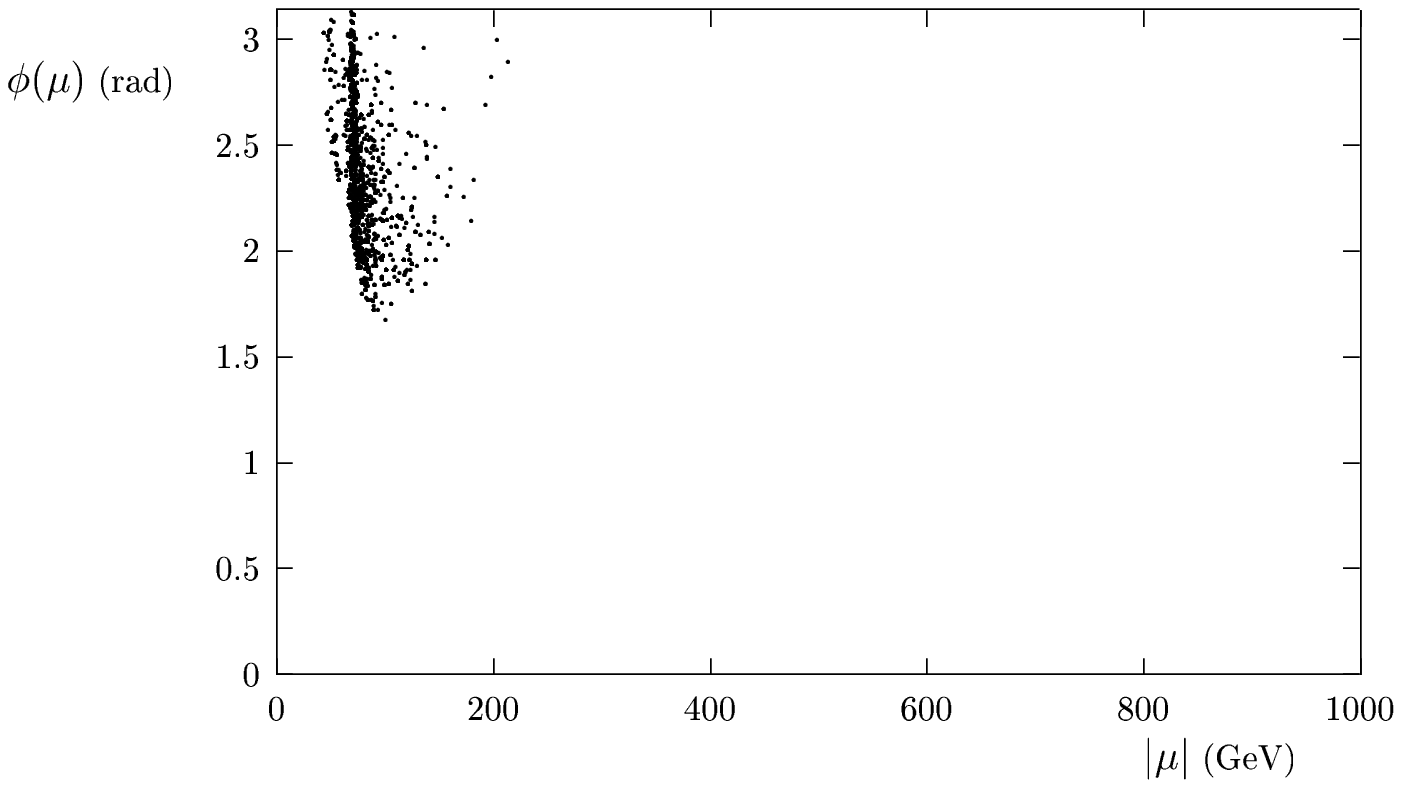}}
\end{picture}
\end{minipage}
\vspace{9cm}
\caption{ \label{fig6_1} 
Correlations in the ($|\mu|$, $\Phi_\mu$) plane from consistency of the
inversion, with incomplete knowledge of the input parameters: 
$M_{\chi^+_1} =$ 80 GeV, 80 $< M_{\chi^+_2} <$ 1000,
 $0 < \phi_L < \pi/2$,  $M_{N_1}=40$, $M_{N_2}=70$;  $1 < \tan\beta <3 $.}


\begin{minipage}[b]{19.cm}
\setlength{\unitlength}{1.in}
\begin{picture}(1.2,1)(0.8,10.3)
\centerline{\epsffile{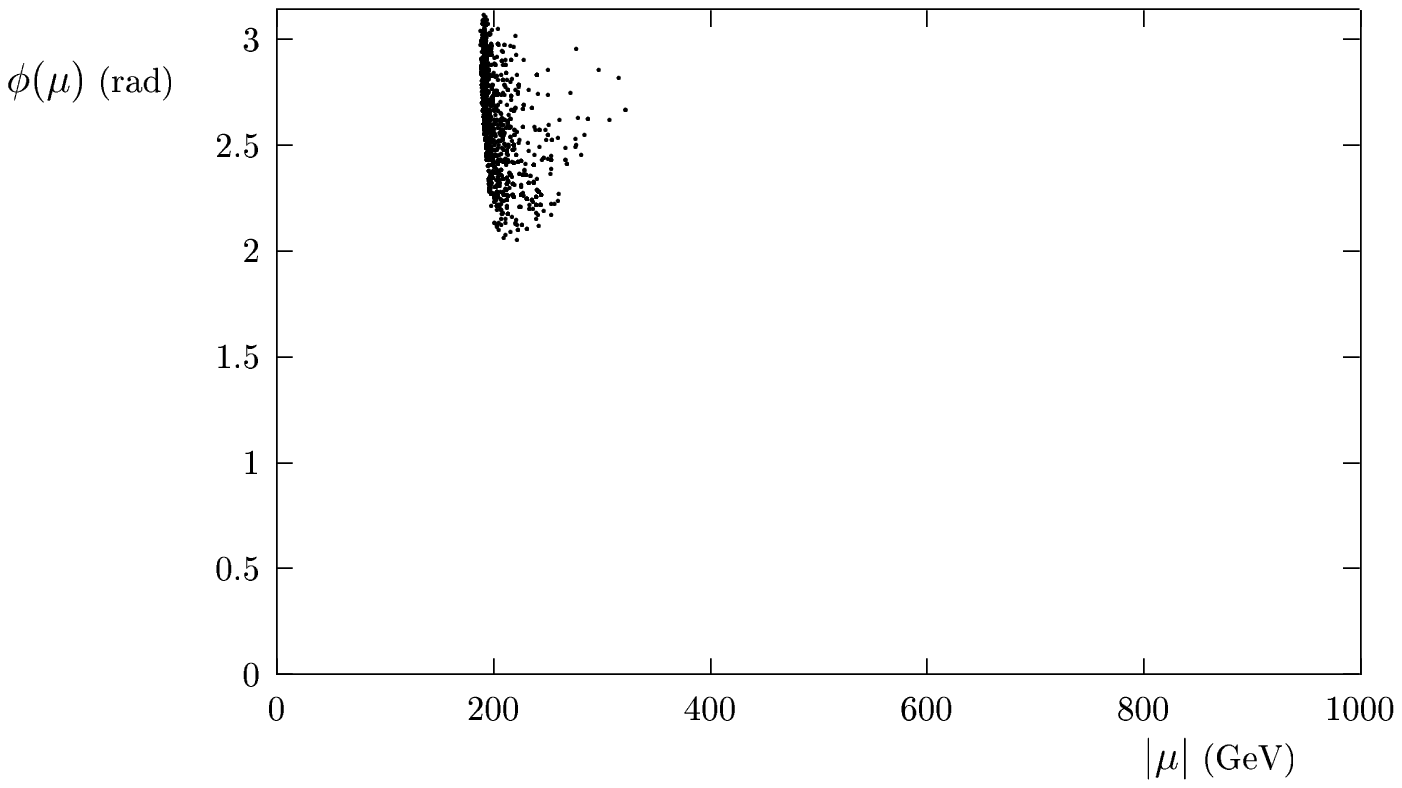}}
\end{picture}
\end{minipage}
\vspace{9cm}
\caption{ \label{fig6_2} same as fig. 7 with a different
input choice: $M_{\chi^+_1} =$ 200 GeV, 200 $< M_{\chi^+_2} <$ 1000, 
$0 < \phi_L < \pi/2$,  $M_{N_1}=150$, $M_{N_2}=190$;  $1 < \tan\beta <3
$.} 
\end{figure}

\newpage
\begin{figure}[htb]
\vspace{-35mm}
\begin{minipage}[t]{19.cm}
\setlength{\unitlength}{1.in}
\begin{picture}(0.2, 0)(0.8,10.3)
\centerline{\epsffile{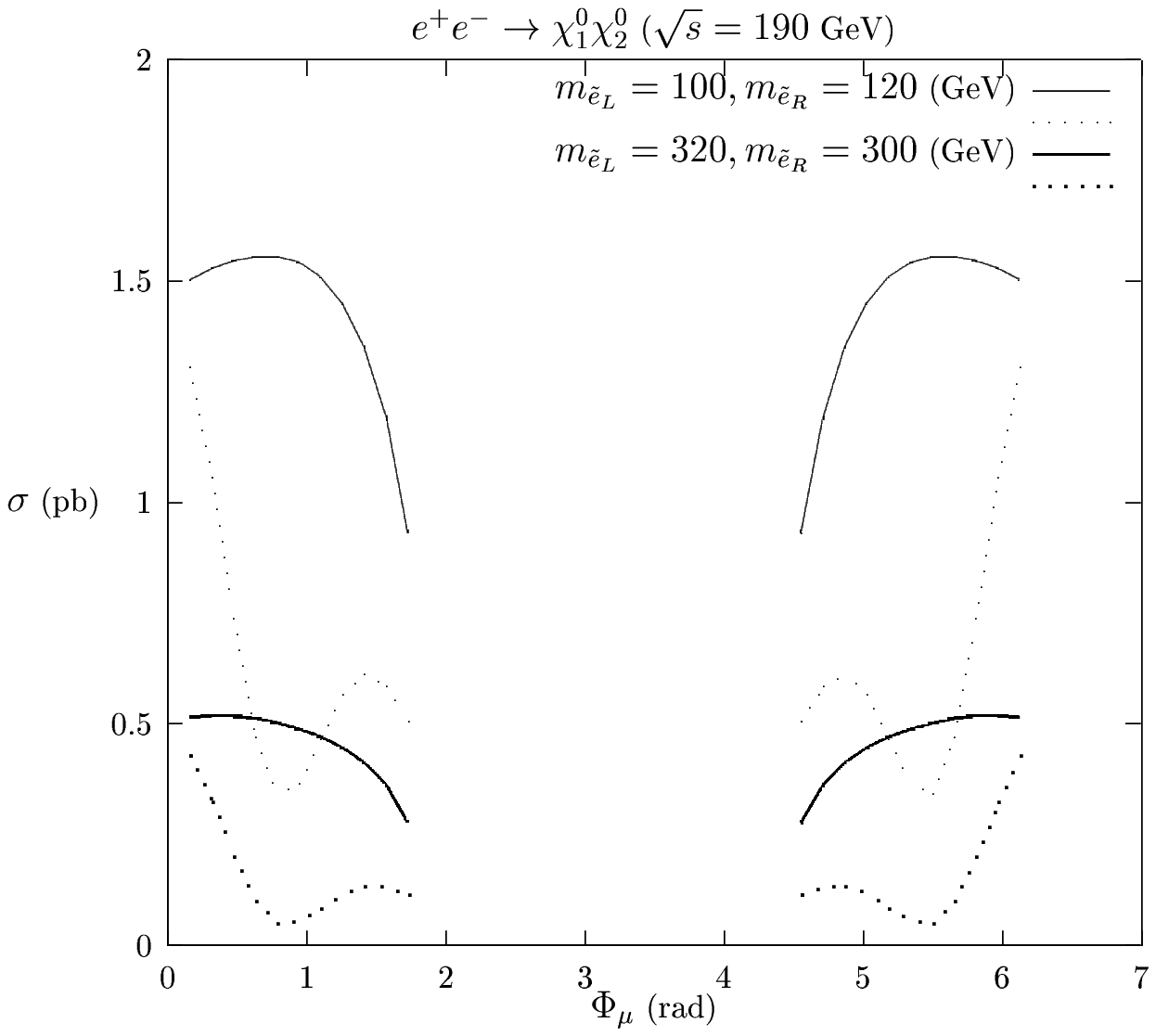}}
\end{picture}
\end{minipage}
 \vskip 9 cm

\begin{minipage}[b]{19.cm}
\setlength{\unitlength}{1.in}
\begin{picture}(1.2,1)(0.8,10.3)
\centerline{\epsffile{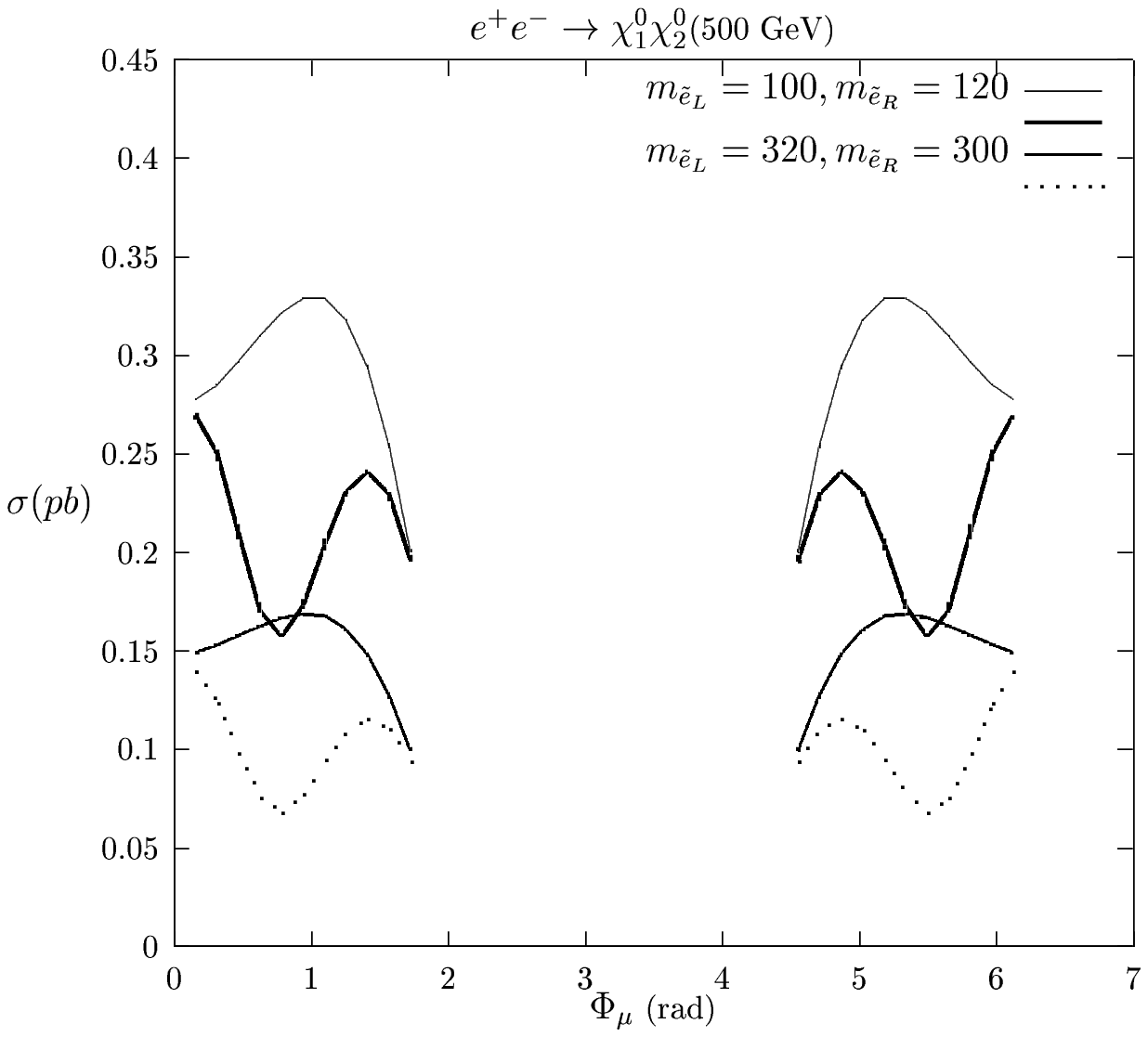}}
\end{picture}
\end{minipage}
\vspace{12cm}
\caption{ \label{nino_fig1} $\chi^0_1, \chi^0_2$ production
cross-section at $\sqrt{s}= 190, 500$ GeV, versus $\Phi_\mu$,
for two different choices of selectron masses, and input choice as in
fig. 1. Plain (dotted) lines correspond to the first (second) $M_1$
solution of fig. 1.}
\end{figure}
\newpage
\begin{figure}[htb]
\vspace{-20mm}
\begin{minipage}[t]{19.cm}
\setlength{\unitlength}{1.in}
\begin{picture}(0.2, 0)(0.8,10.3)
\centerline{\epsffile{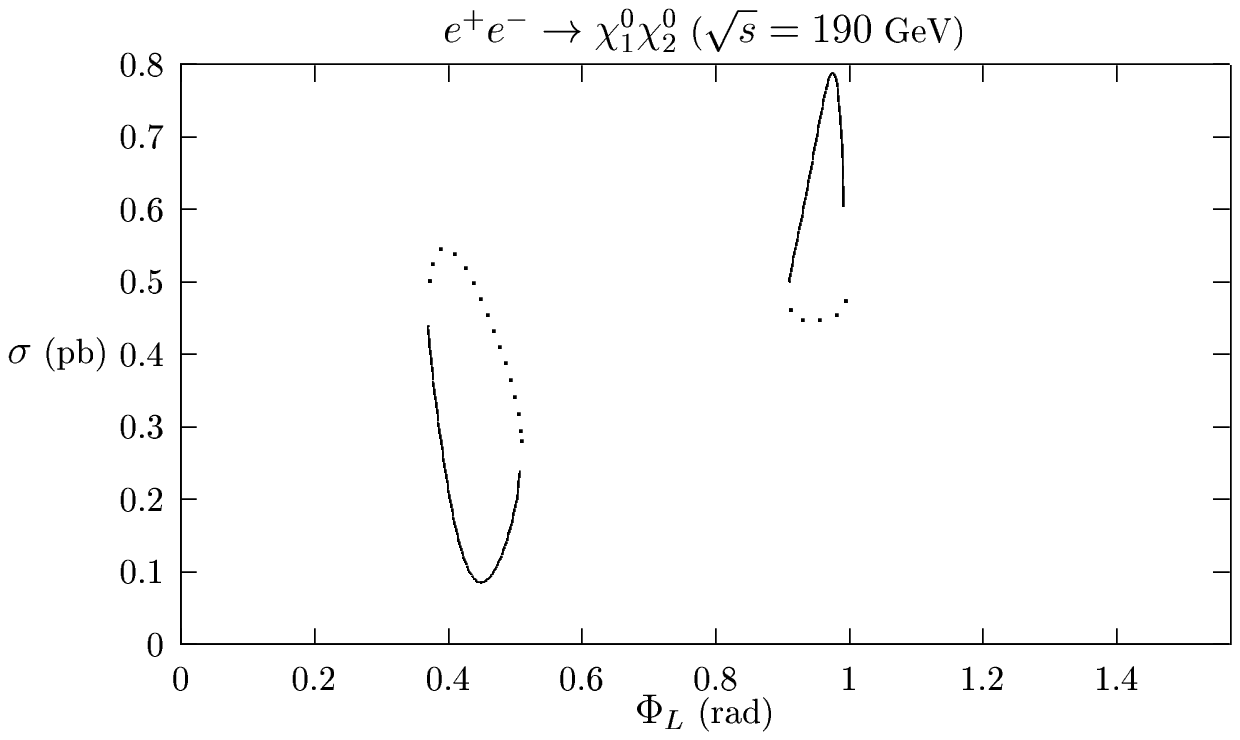}}
\end{picture}
\end{minipage}

\vskip 9 cm

\begin{minipage}[b]{19.cm}
\setlength{\unitlength}{1.in}
\begin{picture}(1.2,1)(0.8,10.3)
\centerline{\epsffile{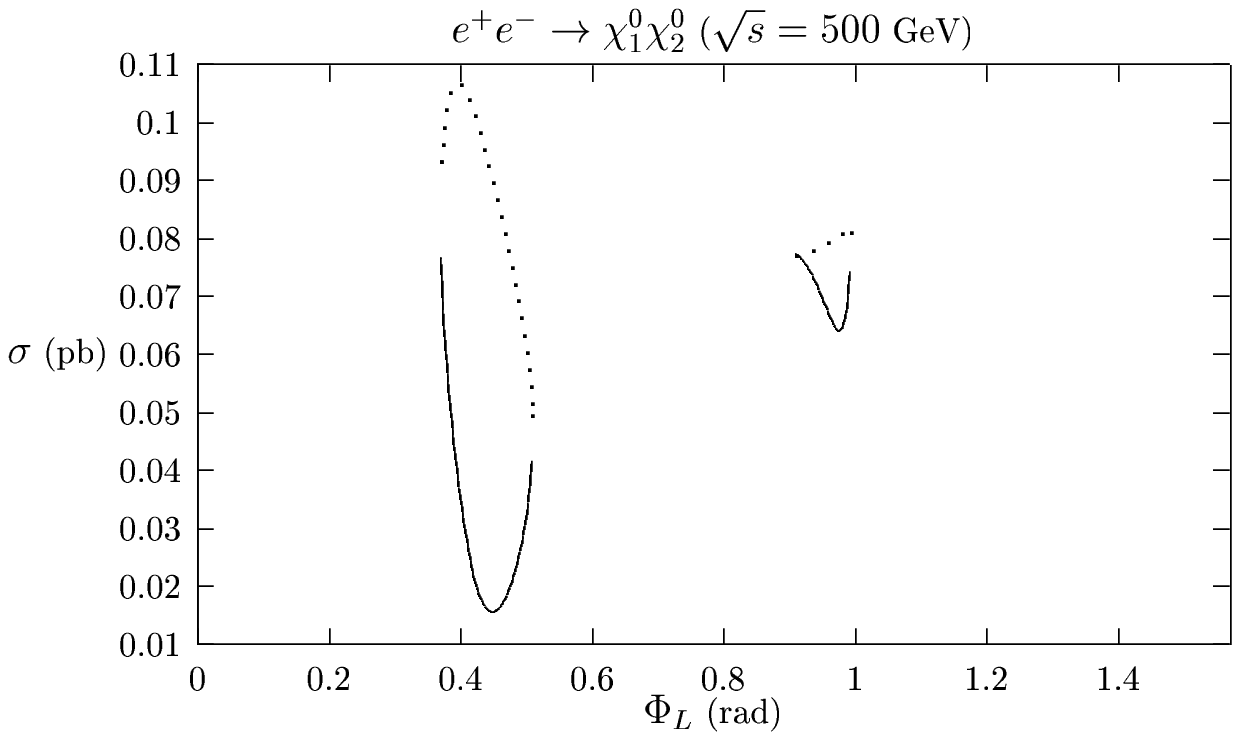}}
\end{picture}
\end{minipage}
\vspace{9cm}
\caption{ \label{nino_fig2} $\chi^0_1, \chi^0_2$ production
cross-section at $\sqrt{s}= 190, 500$ GeV, versus the chargino mixing
angle $\phi_L$, and for the same physical input as fig. 3.
Plain (dotted) lines correspond to the first (second) $M_1$
solution of fig. 3.}
\end{figure}


\newpage

\begin{thebibliography}{99}  

\bibitem{R1} For reviews see: H.P.~Nilles, Phys. Rep. 110, 1 (1984); \\ 
H.E.~Haber and G.L.~Kane, Phys. Rep. 117, 75 (1985); 
\bibitem{martin} see for instance
S.~P. Martin, hep-ph/9709356, in {\em Perspectives in 
Supersymmetry}, ed. by G.~L. Kane, World Scientific.
%
\bibitem{softbreak} L. Girardello and M.T. Grisaru, Nucl.Phys. B194 (1982) 65.
%
\bibitem{CP1} J. Ellis, S. Ferrara and D. Nanopoulos, Phys. Lett. B114
(1982) 231; 
W. Buchm\"uller and D. Wyler, Phys. Lett. B121 (1983) 321;
J. Polchinski and M. Wise, Phys. Lett. B125 (1983) 393;
F. del Aguila, M. Gavela, J. Grifols and A. Mendez, Phys. Lett. B126
(1983) 71; 
D. Nanopoulos and M. Srednicki, Phys. Lett. B128 (1983) 61.
%
\bibitem{CPrev} For recent reviews on CP violation in supersymmetry
and FCNC constraints, see e.g. Y. Grossman, Y. Nir and R. Rattazzi,
hep-ph/9701231, in ``Heavy flavours II" eds. A. Buras and M. Lindner,
World Scientific;\\ A. Masiero and L. Silvestrini, hep-ph/9711401,
Lectures at the Erice school, Sept. 1997.
%
\bibitem{fcncexp} S. Adler et al, Phys. Rev. Lett. 79 (1997) 2204.
%
\bibitem{PDG98} Review of Particle Physics, C.~Caso et al.,
   Eur.~Phys.~J.~C {\bf 3} (1998) 1.
%
\bibitem{FalOl} T. Falk and K.A. Olive, Phys. Lett. B375 (1996)
196; 
Phys. Lett. B439 (1998) 71.
%
%
\bibitem{edme} E. Commins et al, Phys; Rev; A50 (1994) 2960;\\
K. Abdullah et al, Phys. Rev. Lett. 65 (1990) 234.
%
\bibitem{edmn} P.G. Harris et al, Phys. Rev. Lett. 82, 904 (1999).
%
\bibitem{IbNa} T. Ibrahim and P. Nath, Phys. Lett. B418 (1998) 98; 
Phys. Rev. D57 (1998) 478;
Phys. Rev. D58 (1998) 111301.
%
\bibitem{BrKa} M. Brhlik and G.L. Kane, Phys. Lett. B437 (1998) 331;\\ 
M. Brhlik, G.J. Good and G.L. Kane, Phys. Rev. D59 (1999) 115004;
M. Brhlik, L. Everett, G.L. Kane and J. Lykken, hep-ph/9905215.
%
\bibitem{BartlCP} A. Bartl, T. Gajdosik, 
W. Porod, P. Stockinger and H. Stremnitzer, 
hep-ph/9903402.
%
\bibitem{Savoyetal} S. Pokorski, J. Rosiek and C. Savoy, hep-ph/9906206.
%
\bibitem{SNOWMASS} I. Hinchliffe et
al., Phys. Rev. D55 (1997) 5520; CMS Collaboration (S. Abdullin et
al.), hep-ph/9806366; 
\bibitem{Denegri} D. Denegri, W. Majerotto and
L. Rurua, Phys. Rev. D58 (1998) 095010.   %
%
\bibitem{inoinvreal} J.-L. Kneur and G. Moultaka, Phys. Rev. D59 (1999)
015005. 
%
\bibitem{choi1} 
S.Y. Choi, A. Djouadi, H. Dreiner, J. Kalinowski and P.M. Zerwas, 
Eur. Phys. J. C7 (1999) 123.
%
\bibitem{choi2} S.Y. Choi, A. Djouadi, H.S. Song and P.M. Zerwas, 
Eur. Phys. J. C8 (1999) 669; hep-ph/9812236.  
%
\bibitem{bartlold} A. Bartl {\sl et al.} Phys. Rev. D40 (1989) 1594.
%
\bibitem{Hphases} D. Demir, hep-ph/9901389;\\
B. Grzadkowski, J.F. Gunion and J. Kalinowski, hep-ph/9902308.
%
\bibitem{PilWag} A. Pilaftsis and C.E.M. Wagner, hep-ph/9902371.
%
\bibitem{ambrosanio}
S. Ambrosanio and B. Mele, Phys.Rev.D52:3900-3918,1995 
%
\bibitem{LEPYR96} Physics at LEP2, CERN 96-01 vol. 1 p.463, 
eds. G. Altarelli, T. Sj\"ostrand and F. Zwirner.
%
\bibitem{inobound} 
ALEPH Collaboration (R. Barate et al.), CERN-EP-99-014, Feb. 1999;
DELPHI Collaboration (P. Abreu et al.), Phys. Lett B446 (1999) 75;
OPAL Collaboration (G. Abbiendi et al.), Eur. Phys. J. C8 (1999) 255;
L3 Collaboration (M. Acciarri et al.), Eur. Phys. J. C4 (1998) 207; 
CDF collaboration (F. Abe et al.), Phys. Rev. Lett. 80 (1998) 5275.
%
\bibitem{lcrep} 
ECFA/DESY LC Physics Working Group report,
Phys. Rept. 299 (1998) 1, and references therein.
%
\bibitem{n1n2lc}
G. Moortgat-Pick, H. Fraas, A. Bartl and W. Majerotto,
hep-ph/9903220. 
%
\bibitem{GDR-MSSM}   for a  
recent collection of supersymmetric particle LEP 
limits, see e.g. A. Djouadi,
S. Rosier--Lees {\it et al.}, GDR-SUSY MSSM working group 
report, hep--ph/9901246.  
%
\bibitem{Petcov} S.T. Petcov, Phys. Lett. B178 (1986) 57. 
%
\bibitem{een1n2phase} 
N. Oshimo, Z. Phys. C41 (1988) 129; 
Y. Kizukuri and N. Oshimo, Phys. Lett. B249 (1990) 449.
%
\bibitem{bartl1}
A. Bartl, H. Fraas and W. Majerotto, Nucl. Phys. B278 (1986) 1
%
\bibitem{rcmass}
D. Pierce and A. Papadopoulos, Nucl. Phys. B430 (1994) 278.
%
\bibitem{rcsig}
M.A. Diaz, S.F. King and D. A. Ross, Nucl. Phys. B529 (1998) 23. 
%
\end{thebibliography}
\end{document}